\documentclass[fleqn,usenatbib]{mnras}

\usepackage{newtxtext,newtxmath}
\usepackage[T1]{fontenc}
\usepackage{ae,aecompl}

\usepackage{graphicx}	% Including figure files
\usepackage{amsmath}	% Advanced maths commands
\usepackage{amssymb}	% Extra maths symbols
\usepackage[para,online,flushleft]{threeparttable}
\DeclareMathOperator\erf{erf}
\title[Magnetar Rates and Evolution]{Formation Rates and Evolution Histories of Magnetars}
\author[P. Beniamini al.]{
	Paz Beniamini,$^{1,2}$\thanks{E-mail: paz.beniamini@gmail.com}
	Kenta Hotokezaka,$^{3}$
	Alexander van der Horst$^{1,2}$ \newauthor
	and Chryssa Kouveliotou$^{1,2}$
	\\
	$^{1}$Department of Physics, The George Washington University, Washington, DC 20052, USA \\
	$^2$Astronomy, Physics and Statistics Institute of Sciences (APSIS)\\	
	$^3$Department of Astrophysical Sciences, Princeton University, 4 Ivy Lane, Princeton, NJ 08544, USA}
\begin{document}
	\label{firstpage}
	\pagerange{\pageref{firstpage}--\pageref{lastpage}}
	\maketitle
	% Abstract of the paper
	\begin{abstract}
		We constrain the formation rate of Galactic magnetars directly from observations. Combining spin-down rates, magnetic activity, and association with supernova remnants, we put a 2$\sigma$ limit on their Galactic formation rate at $2.3-20\mbox{kyr}^{-1}$. This leads to a fraction $0.4_{-0.28}^{+0.6}$ of neutron stars being born as magnetars. We study evolutionary channels that can account for this rate as well as for the periods, period derivatives and luminosities of the observed population. We find that their typical magnetic fields at birth are $3\times 10^{14}-10^{15}$~G, and that those decay on a time-scale of $\sim 10^4$~years, implying a maximal magnetar period of $P_{\rm max}\approx 13$~s. A sizable fraction of the magnetars' energy is released in outbursts. Giant Flares with $E\geq 10^{46}$ erg are expected to occur in the Galaxy at a rate of $\sim 5\mbox{kyr}^{-1}$. Outside our Galaxy, such flares remain observable by {\it Swift} up to a distance of $\sim 100$~Mpc, implying a detection rate of $\sim 5\mbox{ yr}^{-1}$. The specific form of magnetic energy decay is shown to be strongly tied to the total number of observable magnetars in the Galaxy. A systematic survey searching for magnetars could determine the former and inform physical models of magnetic field decay.
	\end{abstract}
	
	
	%%%%%%%%%%%%%%%%%%%%%%%%%%%%%%%%%%%%%%%%%%%%%%%%%%
	\voffset = -0.5in  
	%%%%%%%%%%%%%%%%% BODY OF PAPER %%%%%%%%%%%%%%%%%%
	\section{Introduction}
	Magnetars represent a rare population of neutron stars with extreme surface magnetic field strengths of $B\gtrsim 10^{14}$~G \citep{Kouveliotou1998}. Observationally they exhibit some unique properties with respect to their  persistent X-ray emission, including their random outbursts, bursts,  and rare Giant Flares \citep[GFs; see, e.g.][for a recent review]{Kaspi2017ARA&A}. \cite{ThompsonDuncan95,Thompson1996} suggested that it is the decay of their magnetic fields which powers both their quiescent emission and outburst behaviour. 
	
	The spin-down life times of normal pulsars, as estimated by their spin-down rates, are much longer than the decay scales of their magnetic fields (depending on the field strengths), which range on a time scale of $\sim 1$--$100$ kyr \citep{Goldreich1992ApJ}. In contrast, the spin-down ages of magnetars can be comparable to  their field decay times. Even though there are currently only 23  confirmed magnetars in the Galaxy, LMC and SMC (compared to $\sim 2700$ pulsars), such short life times suggest that their formation rate may be quite high. Indeed, previous works have estimated the magnetar formation rate at $\sim1-10\,\%$ of all pulsars \citep{Kouveliotou1994,Heyl1998ApJ,Woods,Gill2007MNRAS,Gullon2015}.
	
	Despite the fact that the number of confirmed magnetars has increased in the past decade, their formation paths and birth rates still remain a mystery. A variety of formation scenarios have been proposed in the literature, including strongly magnetized stars \citep{Ferrario2005MNRAS}, very massive stars \citep{Muno2006ApJ}, and a binary origin \citep{Popov2006MNRAS,Bogomazov2009ARep}. The magnetar birth rate is one of the key quantities to unveil their true progenitors. We explore here the constraints on the Galactic formation rate of magnetars directly from observations, and compare them to the formation rates of massive stars and neutron stars. We present estimates for the ages of all observed magnetars based on different observables, including their periods and period derivatives, persistent X-ray luminosities, GF energetics, and possible associations with supernova remnants (SNRs); all these ages can be used to estimate the magnetar formation rates. The results of this analysis are presented and briefly discussed in \S \ref{sec:rates}.
	
	In Section \S \ref{sec:sample} we describe two possible sample selections, those of confirmed and `candidate' magnetars. In \S \ref{sec:model} we present the models describing the spin and magnetic field evolution and describe their impact on the observable parameters. By means of a statistical analysis, comparing simulations with the observed population, we show in \S \ref{sec:like} the most likely ranges for the model parameters, in particular the magnetic field at birth, the magnetic field decay time, and a parameter describing the evolution rate. We compare the results of our simulations with the observed magnetar region on the $P-\dot{P}$ diagram, their outburst activity energetics, and their log$N$-log$S$ distribution. We also discuss the consequences of the inclusion in the Galactic magnetar population, two radio pulsars, which have displayed magnetar-like activity. A summary and conclusions are presented in \S \ref{sec:conclusions}.

	\section{Galactic Magnetar birth rates}
	\label{sec:rates}
	Magnetar ages are usually estimated from their observed period~$P$ and period derivative~$\dot{P}$: $\tau_{\Omega}=\lvert \frac{P}{2\dot{P}} \rvert$, which can be written as $\tau_{\Omega}=\lvert \frac{P}{(n-1)\dot{P}} \rvert$ to include the more general case of a braking index $n\neq 3$. Since magnetars often exhibit variability in their $\dot{P}$ evolution, this can result in a range of possible life-times for a given system, $\tau_{\Omega,\rm min}-\tau_{\Omega,\rm max}$.
	
	As the rotational energy loss of magnetars is insufficient to power their X-ray luminosity, it is thought that the latter is instead powered by magnetic energy losses \citep{Thompson1996,Perna2011,TB2007,Beloborodov2013}. This then leads to another important time-scale that may determine a magnetar's age, which is the decay time of the magnetic field, defined here as $\tau_B\equiv \lvert \frac{B}{\dot{B}} \rvert$. Note that the actual time scale of field decay can be altered from $\tau_B$ by a constant factor of order unity. Since the relation between $B$ and $\dot{B}$ is not directly constrained from observations, we do not include this constant in the definition of $\tau_B$. We revisit the estimate of $\tau_B$ under different  magnetic field evolutionary paths in \S \ref{sec:model}.
	
	For any spin-down and magnetic field decay models, the real age of the magnetar is dictated by the shorter of the two time scales: $\tau=\min(\tau_{\Omega},\tau_B)$. The reason is that once the magnetic field decays significantly, the spin-down strongly decreases, and the measured $\tau_{\Omega}$ will become larger than the real age. Observationally, $\tau_B$ can be constrained from the magnetar persistent X-ray luminosity, $L_{\rm p}$, and their magnetic energy content $E_B$, that can be estimated assuming a dipolar field from the observed $P$ and $\dot{P}$,  as:
	\begin{equation}
	\tau_{\rm B,p}\approx \bigg \lvert \frac{E_B(P,\dot{P})}{\dot{E}_B}  \bigg \rvert \lesssim  \bigg \lvert \frac{E_B(P,\dot{P})}{L_{\rm p}} \bigg  \rvert.
	\end{equation}
	If the total energy output of magnetars is instead dominated by wavelengths beyond the X-ray band, the real age would be lower than the estimate given above (thus rendering our limits on the rates below conservative). Furthermore, additional energy loss in the form of bursts in outbursts would also decrease the real age compared to $\tau_{\rm B,p}$. We do not attempt to estimate this contribution here. We do, however, track the energy loss by GFs, which can indeed result in huge energy losses. This was demonstrated in the GF of SGR 1806-20, which released over $10^{46}\mbox{erg}$ \citep{Hurley05}, a fraction of $\sim 0.08-0.19$ of its magnetic energy (assuming a dipole field) in one event. Given that only 3 GFs have been observed so far, from 3 different magnetars, the main uncertainty for the GFs' contribution to the magnetic field decay time is their intrinsic rate.

	A final estimate of magnetars' life times can be obtained in those cases where the magnetars are associated with SNRs, using the size of the SNR and their plasma temperature. 
	
	In Table \ref{tbl:ages} we present the age upper limits, or estimates in case of a SNR association, of all known magnetars and two magnetar candidates (PSR~J1119-6127 and PSR~J1846-0258, which are classified as pulsars but have exhibited magnetar-like activity) for which $P,\dot{P},L_p$ are measured. For a number of systems there is some uncertainty in evaluating the time scales. For example, this can occur due to $\dot{P}$ and $L_p$ having changed over the observed period, resulting in a range of $\tau_{\Omega}$ and $\tau_B$ values. In those cases we quote a range of values $\tau_{\rm X,L}-\tau_{\rm X,U}$, corresponding to the full range of uncertainty and/or evolution of the observables in that system. 
	
	Taking only confirmed magnetars residing in the Galaxy, SMC and LMC, we can estimate a $2\sigma$ lower limit on the Galactic birth rate of magnetars. This is done by assuming a Poisson distribution for the number of magnetars born per year characterized by a rate $\Phi_X$ and finding the ranges of $\Phi_X$ for which at least $0.05$ of realizations would violate the observed distribution of age upper limits $\tau_X$. For the case of the SNR age estimates, an upper limit on the rate can also be calculated in an equivalent way.
	
	The resulting $2\sigma$ rates are (numbers in parentheses include also the two magnetar candidates):
	\begin{eqnarray}
	\label{eq:rates}
	& \Phi_{\Omega,{\rm L}}>1.4 \ (>1.9) \mbox{ ; } \Phi_{\Omega,{\rm U}}>2.6 \ (>3.2)\mbox{ kyr}^{-1} \nonumber \\
	& \Phi_{\rm B,p,L}>0.5 \ (>3.4) \mbox{ ; } \Phi_{\rm B,p,U}>0.6 \ (>9.4) \mbox{ kyr}^{-1} \nonumber \\
	& \Phi_{\rm min}>2 \ (>10) \mbox{ ; } \Phi_{\rm max}>3.7 \ (>10) \mbox{ kyr}^{-1} \nonumber \\
	& \Phi_{\rm SNR}= 0.6-2.5 \ (0.6-2.8) \mbox{ kyr}^{-1}
	\end{eqnarray}
	
	where $\Phi_{\rm X,L}$ ($\Phi_{\rm X,U}$) is the $2\sigma$ lower limit on the rate associated with $\tau_{\rm X,L}$ ($\tau_{\rm X,U}$). $\Phi_{\rm min}$ ($\Phi_{\rm max}$) is the rate associated with $\tau=\min(\tau_{\rm \Omega,L},\tau_{\rm B,L})$ ($\tau=\min(\tau_{\rm \Omega,U},\tau_{\rm B,U})$). We combine the lower limits from $\Phi_{\rm min}$ and the estimates from $\Phi_{\rm SNR}$ to improve our estimates of the formation rate. There are two confirmed magnetars, SGR 0526-66 and SGR 1627-41 for which the estimated age from the SNR association is higher than (and therefore in contradiction with) the upper limit from spin-down age. In both cases, however, the discrepancy is by a factor of $\lesssim 3$. Such a discrepancy may arise from a change in the braking index or uncertainties in the SNR modeling. We conservatively take the age as estimated by the SNR for those two cases (this will decrease our estimated formation rates). Finally, taking into account all these lifetime estimates, we derive a confirmed magnetar rate of
	\begin{eqnarray}
	\Phi_{\rm mag}=2.3-20\,\mbox{kyr}^{-1},
	\end{eqnarray}
	where the quoted range is at a confidence level of 2$\sigma$. We address the potential contribution of the two magnetar candidates to the formation rates in \S \ref{sec:magcand}.

	\begin{table*}
		% 		\footnotesizea
		\begin{center}
			\caption{Age estimates for different magnetars (magnetar candidates shown in parentheses).}
			\begin{threeparttable}
				\begin{tabular}{ccccccc}\hline		
					Name &  $\tau_{\Omega}$  \tnote{(a)} & $\tau_{\rm B,p}$ \tnote{(b)}& $\tau_{\rm B,GF}$ \tnote{(c)}& $\tau_{\rm SNR}$\tnote{(d)} & comment & refs \tnote{(e)}. \\ \hline
					CXOU J010043.1-721134 & 6.8 &  14  & - & - & SMC & 1\\
					4U 0142+61 & 68 & 0.3& - & - & - & 2,41 \\
					SGR 0418+5729 & $3.6\times 10^4$ & 220 & - & - & - & 3\\
					SGR 0501+4516 & 15 & 99& - & 4-7 & SNR (G160.9+02.6) & 4,5,45\\
					SGR 0526-66 & 1.9-3.4 & 10-17& 26-44& 4.8 & LMC & 6,7\\
					1E 1048.1-5937 & 2-8 & 10-40 & - & - & - & 8,9 \\
					(PSR J1119-6127) & 1.6 & 0.02& - & 4.2-7.1 & SNR (G292.2-0.5) & 10,11,12,46\\
					1E 1547.0-5408 & 0.7-1.4 & 220-460 & - & - & SNR (G327.24-0.13) & 13,14,15 \\
					PSR J1622-4950 & 3.5-7.3 & 550-1100 & - &$<6$ & SNR (G333.9+00.0)  & 16 \\
					SGR 1627-41 & 2.2 & 83 & - & 5 & SNR (G337.0-00.1) & 17,18 \\
					CXOU J164710.2-455216 & $>173$ & $<138$ & - & - & Westerlund & 19,20 \\
					1RXS J170849.0-400910 & 9 & 9.3& - & - & - & 2,42,43 \\
					CXOU J171405.7-381031 & 0.58-1 & 24-43 & - & 0.35-3.15 & SNR (CTB 37B) & 21,22,23 \\
					SGR J1745-2900 & 4.3-9.7 & $>2800$ & - & - & Galactic center  & 24 \\
					SGR 1806-20 & 0.16-1.45 & 6.1-55 & 0.6-5.3 & - & W31 & 25,26 \\
					XTE J1810-197& 8.4-20 & 3000-8000 & -& - & - & 13 \\
					Swift J1822.3-1606& 1600-6400 & $>2600$ & - & - & - & 27,28 \\
					SGR 1833-0832& 34 & - & - & - & - & 29 \\
					Swift J1834.9-0846  & 4.9 & $>1.4\times 10^4$ & - & - & - & 30 \\
					1E 1841-045  & 4.5 & 5.9 & - & 0.75-2.1  & SNR (Kes 73) & 2,31,44 \\
					(PSR J1846-0258) & 0.73 & 0.05& - & 0.35-0.56 & SNR (Kes 75) & 32,33,34,46 \\
					3XMM J185246.6+003317 & $>1300$ & - & - & 4.4-6.7 & SNR (G033.6+00.1) & 35,36 \\
					SGR 1900+14 & 0.46-1.3 & 8.8-25 & 30-83& - & - & 26,37,38,39 \\
					SGR 1935+2154 & 3.6 & - & - & 41 & SNR (G57.2+0.8) & 40 \\
					1E 2259+586  & 230 & 1.2 & - &   9 & SNR (CTB 109) & 2 \\
					\hline  
					\label{tbl:ages}
				\end{tabular}
				\begin{tablenotes}
					\item[(a)] Dipole spin-down age (kyr).
					\item[(b)] Magnetic field decay age from persistent emission (kyr).
					\item [(c)] Magnetic field decay age from GF emission, assuming one GF every 50 years (kyr).
					\item [(d)] SNR age estimate (kyr).
					\item [(e)] References (as ordered in table):  \cite{McGarry2005,Dib2014,Rea2013,Camero2014,Leahy2007A&A,Kulkarni2003,Tiengo2009,Gavriil2004,Dib2009,Camilo2000,Weltevrede2011MNRAS,Gogus2016ApJ,Camilo2007ApJ,Camilo2008,Dib2012,Levin2012,EBP2009,ETM2009,An2013,Rodriguez2014,Sato2010,Halpern2010,Aharonian2008A&A,Kaspi2014,Woods2007,Nakagawa2009,Scholz2014,Rea2012,Esposito2011,Kargaltsev2012,Vasisht1997ApJ,Livingstone2011,Gotthelf2000ApJ,Gavriil1802,Rea2014,Zhou2014ApJ,Marsden1999,Woods2003,Mereghetti2006,Israel2016,denHartog2008,denHartog2008b,Kuiper2006,Kuiper2004,Enoto2010,Parent2011}
				\end{tablenotes}
			\end{threeparttable}
		\end{center}
	\end{table*}   
	\voffset = -0.3in  
	\subsection{Comparison with massive star formation rate}
	To understand the implications of the estimated magnetar formation rate, it is useful to consider the formation rate of massive stars and neutron stars in general. This can be done in various ways. First, the current rate of star formation in the Milky Way is approximately $1.65\pm 0.19 M_{\odot}\mbox{ yr}^{-1}$ \citep{Licquia2015}.
	Assuming a Kroupa initial mass function \citep{Kroupa2001}, this corresponds to a birth rate of stars with $M>8M_{\odot}$ (the majority of which become neutron stars) of roughly $\Phi_{\rm IMF}=17\pm 2\mbox{ kyr}^{-1}$. This is a good approximation of the Galactic neutron star formation rate.

	Next, we can establish the equivalent value of $\Phi_{\Omega}$ (rate from spin-down ages) for all pulsars ($\Phi_{\Omega,\rm pul}$). Using the data reported in the ATNF catalog \footnote{\url{http://www.atnf.csiro.au/research/pulsar/psrcat/}} \citep{ATNF}, we find $\Phi_{\Omega,\rm pul}>3.8\mbox{ kyr}^{-1}$. Surprisingly, this is not much larger than the value for magnetars alone. However, since regular pulsars are detected by their beamed radio emission, there are approximately $P/2W$ actual pulsars in the Galaxy for any observed pulsar with a period $P$ and beam width $W$ \citep{Lyne1988}. Correcting the spin-down age by these weights, and using $W_{50}$ as the beam width (i.e. the width of the pulsar's pulses at a $50\%$ level), we find $\Phi_{\Omega,\rm pul}>14\mbox{ kyr}^{-1}$, in accord with the estimate based on the star formation rate.
	%$\Phi_{\Omega,\rm pul}>5.3\mbox{ kyr}^{-1}$ if I use W10
	
	A final estimate of the massive star formation rate comes from the distribution of SNR ages in the Galaxy. We collect the age estimates of all SNRs clearly associated with core-collapse SNe from the SNR catalog \footnote{\url{http://www.physics.umanitoba.ca/snr/SNRcat/}} \citep{Ferrand2012}. Using those ages we estimate the formation rate of such SNe as $\Phi_{\rm cc-SNe}=8-37\mbox{ kyr}^{-1}$ at a $2\sigma$ confidence level. This range is consistent with the other estimates above.

\begin{figure}
		\centering
		\includegraphics[width=0.48\textwidth]{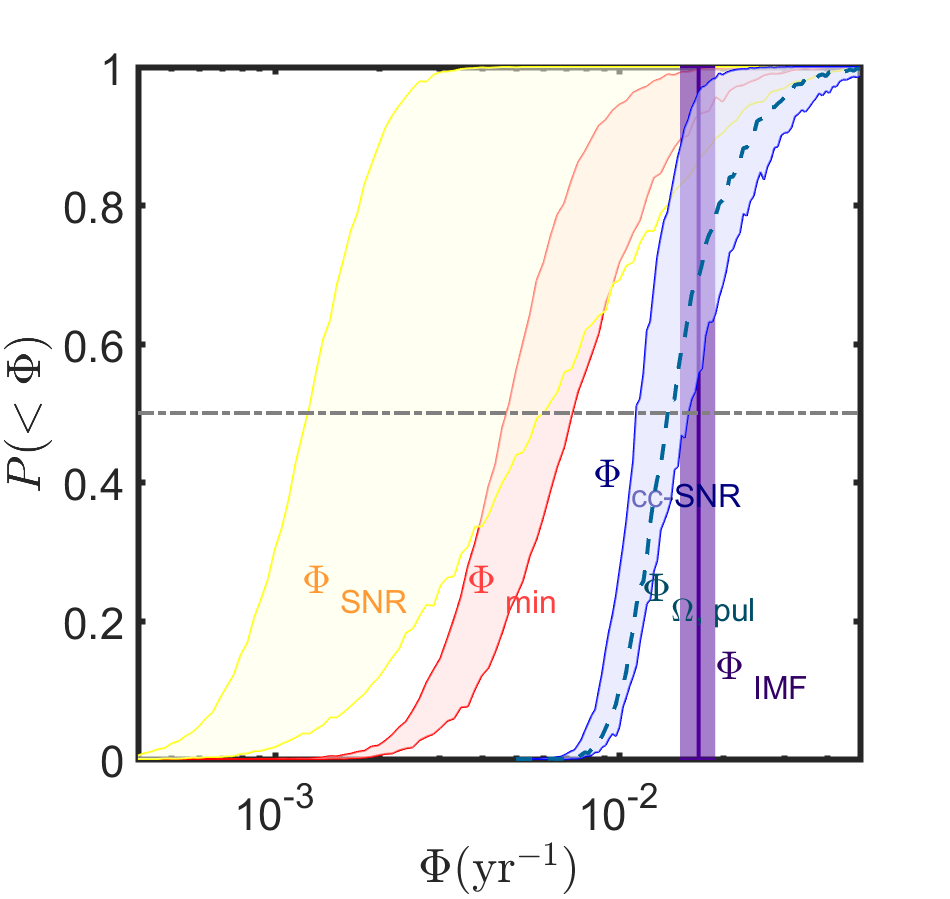}
		\caption{Cumulative probability for the formation rate estimates and lower limits of magnetars as compared to those of massive stars. From left to right: the yellow region depicts the magnetar formation rate estimates based on the SNRs associated with confirmed magnetars ($\Phi_{\rm SNR}$); the red region depicts the lower limit on those rates based on magnetars' spin-down and magnetic ages ($\Phi_{\rm min}$); the blue region is the estimate of Galactic core-collapse SNe based on Galactic SNR ages ($\Phi_{\rm cc-SNR}$); the dashed line is the lower limit on the overall pulsar formation rate based on the spin-down ages of all Galactic pulsars ($\Phi_{\Omega, \rm pul}$); and the purple region is the estimate of massive star formation rate based on the Galactic IMF ($\Phi_{\rm IMF}$). The dot-dashed gray line depicts the median probability for all these curves. }
		\label{fig:ratesummary}
	\end{figure}

	In Figure \ref{fig:ratesummary} we show the different estimates for the formation rates of magnetars and massive stars. Using our best estimates for both rates we conclude that a fraction of $0.4_{-0.28}^{+0.6}$ of neutron stars are born as magnetars (i.e., $12-100\%$ at the $2\sigma$ confidence level). In the next sections we explore different formation and evolution scenarios for the magnetar population.

	\section{Sample selection}
	\label{sec:sample}
	
\begin{figure}
		\centering
		\includegraphics[width=0.5\textwidth]{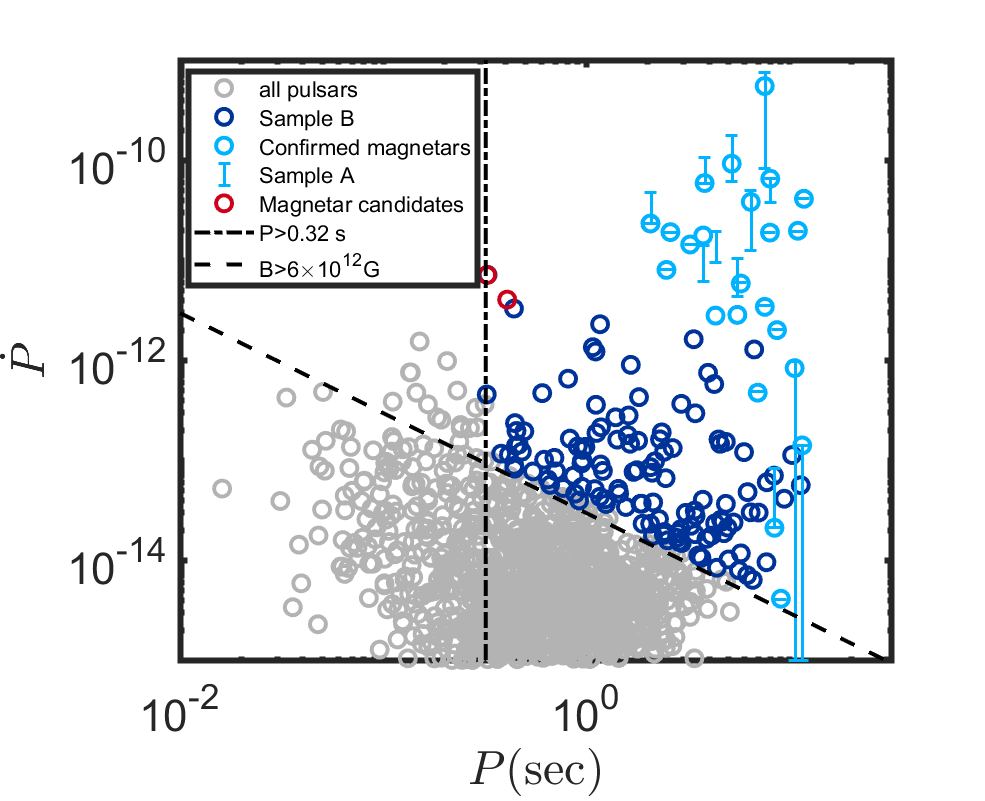}
		\caption{$P-\dot{P}$ diagram showing pulsars (grey), magnetars (also our sample A, depicted in cyan), and the two magnetar candidates (red). A dot dashed (dashed) line denotes the limit $P=0.32$~s ($B_{\rm dip}=6\times10^{12}G$). Systems to the right and top of these lines have spin periods and magnetic fields at least as large as those of the confirmed magnetars and magnetar candidates, which compose our sample B.}
		\label{fig:ppdot}
	\end{figure}

	Figure \ref{fig:ppdot} presents the magnetar population in the $P-\dot{P}$ diagram using magnetar data from the Mcgill catalog~\footnote{\url{http://www.physics.mcgill.ca/~pulsar/magnetar/main.html}} \citep{Olausen2014} and pulsar population data from the ATNF catalog. We denote the two magnetar candidates PSR J1119-6127 and PSR J1846-0258 in red circles. The latter have led us to define two sample selections:
	\begin{itemize}
		\item Sample A: Only confirmed magnetars. This sample consists of the 23 confirmed magnetars in the Milky Way, SMC and LMC for which the period and period derivative are well measured.
		\item Sample B: Motivated by the location of the magnetar candidates in the $P-\dot{P}$ diagram, and given that they have been shown to exhibit magnetar-like behaviours, we consider all pulsars with a spin period and dipole magnetic field at least as large as those of the confirmed magnetars and the two magnetar candidates. This sample consists of 144 neutron stars.
	\end{itemize}
	
	\section{Spin and magnetic field evolutions}
	\label{sec:model}
	For a general model of spin-down, $\dot{\Omega}\propto B^2 \Omega^n$, and a magnetic field decay evolution given by $\dot{B} \propto B^{1+\alpha}$ \citep{Colpi2000}, the magnetic field evolves as:
	\begin{eqnarray}
	\label{eq:Bevolve}
	&    B=B_0\bigg(1+\frac{\alpha t}{\tau_B}\bigg)^{-1/\alpha}\mbox{ for }\alpha\neq 0 \nonumber\\
	& B=B_0\exp(-t/\tau_B)\mbox{ for }\alpha= 0
	\end{eqnarray}
	while the spin evolves with time as:
	\begin{eqnarray}
	\label{eq:spimevolve}
	&    \Omega=\Omega_0 \bigg( \frac{(n-1)\tau_B}{2(\alpha-2)\tau_{\Omega}}\bigg((1+\frac{\alpha t}{\tau_B})^{\alpha-2 \over \alpha}-1\bigg)+1\bigg)^{1\over 1-n}\mbox{ for }n\neq 1,\alpha\neq 0 \nonumber\\
	& \Omega=\Omega_0 \bigg(1+\frac{(n-1)\tau_B}{4\tau_{\Omega}}(1-\exp{(-2t/\tau_B)})\bigg)^{1\over 1-n}\mbox{ for }n\neq 1,\alpha= 0
	\nonumber\\
	& \Omega=\Omega_0 \exp\bigg( -\frac{\tau_B}{2(\alpha-2)\tau_{\Omega}}\bigg((1+\frac{\alpha t}{\tau_B})^{\alpha-2 \over \alpha}-1\bigg)\bigg)\mbox{ for }n=1,\alpha\neq 0
	\nonumber\\
	& \Omega=\Omega_0 \exp\bigg( -\frac{\tau_B}{4\tau_{\Omega}}(1-\exp{(-2t/\tau_B)})\bigg)\mbox{ for }n=1,\alpha= 0
	\end{eqnarray}
	where the subscript 0 denotes a quantity measured at the time of birth of the neutron star. Since $\dot{P}\propto\dot{\Omega}/\Omega^2$, $n<2$ implies $\dot{P}$ increases with time for a constant magnetic field. $n<1$ ($\alpha<0$) implies that the spin (magnetic field) goes to zero at a finite time $t=\tau_{\Omega}$ ($t=\tau_B$). These models may still be physically appropriate for models with fast spin, or super-exponential magnetic field evolution, at times that are sufficiently shorter than the critical times given above.
	Note that for the dipole case, $n=3$ and $\tau_{\Omega}$ is given by
	\begin{equation}
	\tau_{\Omega}=\bigg \lvert \frac{\Omega_0}{2\dot{\Omega}_0} \bigg \rvert=\frac{3c^3I}{4B_0^2R^6\Omega_0^2}
	\end{equation}
	where $R$ is the radius of the neutron star, $I$ is its moment of inertia, and $c$ is the speed of light. Equation \ref{eq:spimevolve} implies that for $\alpha<2$, due to the decay of the magnetic field, the spin-down diminishes over time and the magnetar's spin eventually approaches a minimum value $\Omega_{\rm min}$ (or equivalently the magnetar approaches a maximum period, $P_{\rm max}$),
	\begin{equation}
	\Omega_{\rm min}=\Omega_0\bigg(1+\frac{(n-1)\tau_B}{2(2-\alpha)\tau_{\Omega}}\bigg)^{1\over 1-n}
	\end{equation}
	where we have assumed $\alpha \neq 0$ and $n\neq1$ for clarity. For a dipole evolution with $\tau_B>(2-\alpha)\tau_{\Omega}$ (such that there is an appreciable decay of the spin) this becomes
	\begin{equation}
	\Omega_{\rm min,dip}\approx \sqrt{\frac{3c^3I(2-\alpha)}{4\tau_B B_0^2 R^6}}\approx 0.5\bigg(\frac{10^{15}G}{B_0}\bigg)\bigg(\frac{10^4\mbox{ yr}}{\tau_B}\bigg)^{1/2} \mbox{rad s}^{-1}.
	\end{equation}
	
	Another possibility for the magnetic field and spin evolution is that the field is initially buried deep under the surface of the neutron star, and only emerges to the surface at later times, before starting to decay again. We refer to this case as magnetic field emergence and take the magnetic field evolution to be
	\begin{equation}
	\label{eq:revive}
	B=B_0\exp{\bigg(-\frac{(t-t_e)^2}{2\tau_B^2}\bigg)}
	\end{equation}
	leading to (for $n\neq 1$):
	\begin{equation}
	\Omega=\Omega_0\bigg(1+\frac{(n-1)\tau_B\pi^{1/2}}{4\tau_{\Omega}}\bigg( \erf{\bigg[\frac{t-t_e}{\tau_B}\bigg]}+\erf{\bigg[\frac{t_e}{\tau_B}\bigg]} \bigg)\bigg)^{1\over n-1}
	\end{equation}
	where $t_e$ is the emergence time, the typical time at which the field emerges to the surface and $\erf{[x]}$ is the error function. As before, $\tau_B$, denotes the time scale for significant changes in the surface magnetic field strength (either upwards or downwards).
	Note that the number of parameters required to explain the field evolution in this case is the same as for the case of field decay discussed above (since we now have $t_e$ instead of $\alpha$).

\begin{figure}
		\centering
		\includegraphics[width=0.5\textwidth]{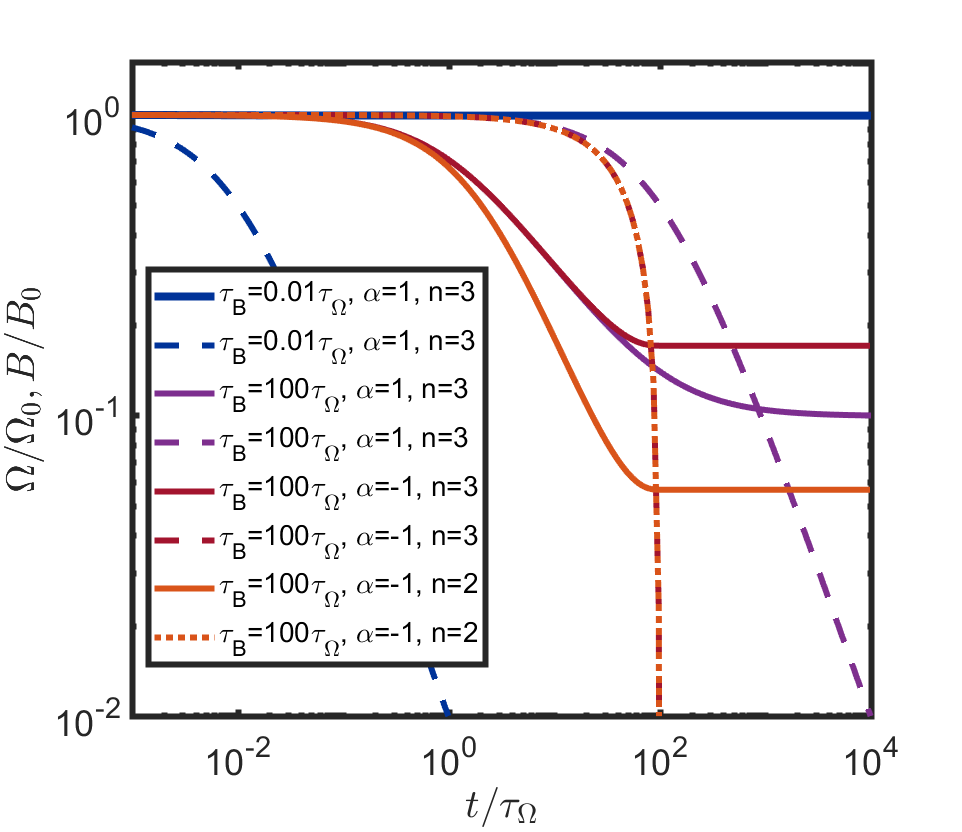}\\
		\includegraphics[width=0.5\textwidth]{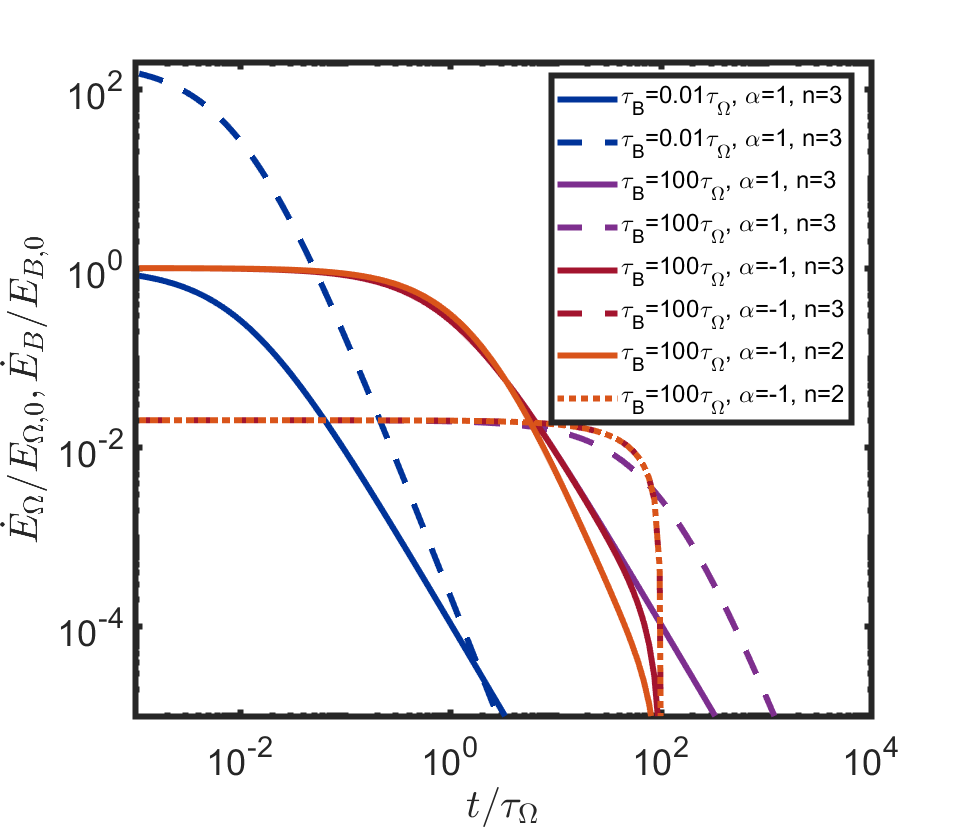}
		\caption{Top: Evolution of spin frequency, $\Omega(t)/\Omega_0$ (solid lines) and surface magnetic field values $B(t)/B_0$ (dashed lines), for the different evolution models discussed in \S \ref{sec:model}. Bottom: Evolution of rotational energy loss rate $\dot{E}_{\Omega}(t)/E_{\Omega,0}$ and magnetic energy loss rate $\dot{E}_{B}(t)/E_{B,0}$ for the same models.}
		\label{fig:Omegat}
	\end{figure}

\begin{figure}
		\centering
		\includegraphics[width=0.5\textwidth]{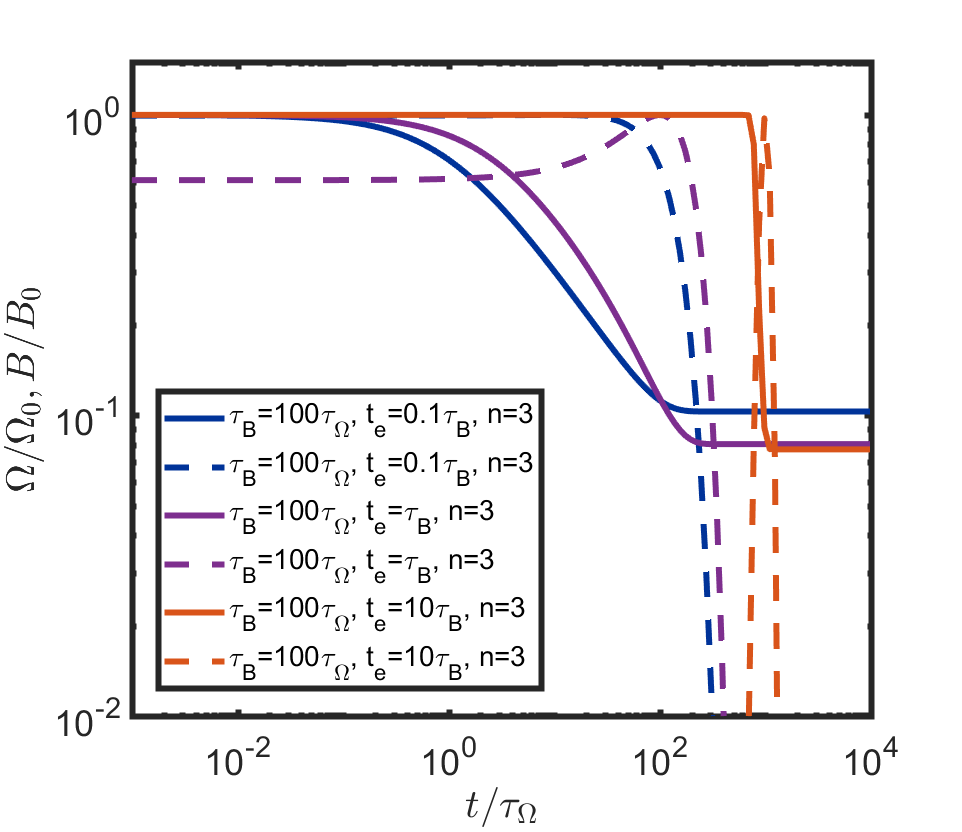}\\
		\includegraphics[width=0.5\textwidth]{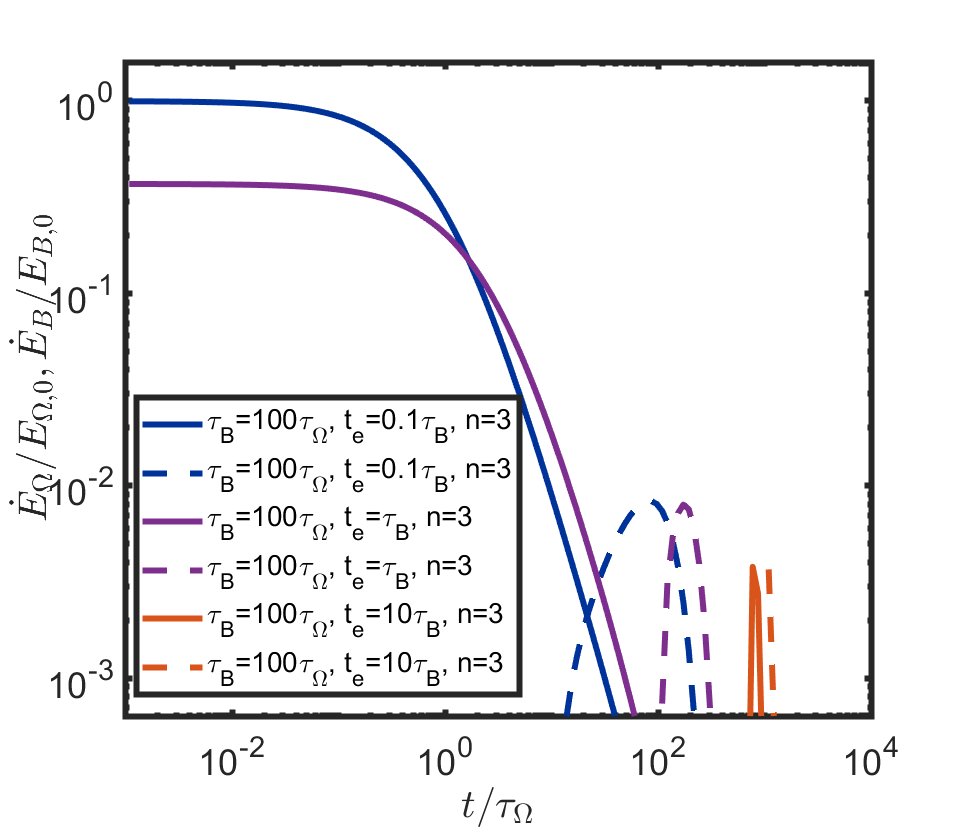}
		\caption{Same as Figure \ref{fig:Omegat}, but for magnetic field emergence models.}
		\label{fig:Omegat2}
	\end{figure}

	Figures \ref{fig:Omegat} and \ref{fig:Omegat2} provide a demonstration of spin and magnetic field evolution for different model parameters, together with the normalized rotational losses
	\begin{equation}
	\frac{\dot{E}_{\Omega}(t)}{E_{\Omega,0}}=\frac{2\Omega \dot{\Omega}}{\Omega_0^2}
	\end{equation}
	and magnetic energy losses
	\begin{equation}
	\frac{\dot{E}_{B}(t)}{E_{B,0}}=\frac{2B \dot{B}}{B_0^2}.
	\end{equation}
	A defining feature of magnetars is that their magnetic energy losses dominate over their rotational energy losses.
	This is determined by the following ratio (which is generally a function of time)
	\begin{equation}
	X_{\dot{E}}(t)\equiv\frac{\dot{E}_{\Omega}}{\dot{E}_B}=\frac{E_{\Omega,0}\tau_B}{2E_{B,0}\tau_{\Omega}}\bigg(\frac{\Omega(t)}{\Omega_0}\bigg)^{n+1} \bigg(\frac{B(t)}{B_0}\bigg)^{\alpha}.
	\end{equation}
	We explore the conditions that lead to a magnetic dominance ($X_{\dot{E}}(t)<1$) for different evolution models below.
	
	Equation \ref{eq:spimevolve} reproduces the intuitive result discussed in \S \ref{sec:rates} that the real age of the magnetar is well estimated by $\tau=\min(\tau_{\Omega},\tau_B)$. Note that in the case where $\tau_B \ll \tau_{\Omega}$, the spin stops decreasing at $\tau_{B}$, but the field vanishes, implying no magnetic or rotational energy release and therefore a system that would no longer exhibit magnetar behaviour. In the case of magnetic field emergence, the rate may actually be shorter than inferred from $\Phi_{\Omega}$, and instead can become dominated by $\tau_B>\tau_{\Omega,0}$. However, as we discuss below, these models are strongly disfavored by observations.
	
	\begin{figure}
		\centering
		\includegraphics[width=0.5\textwidth]{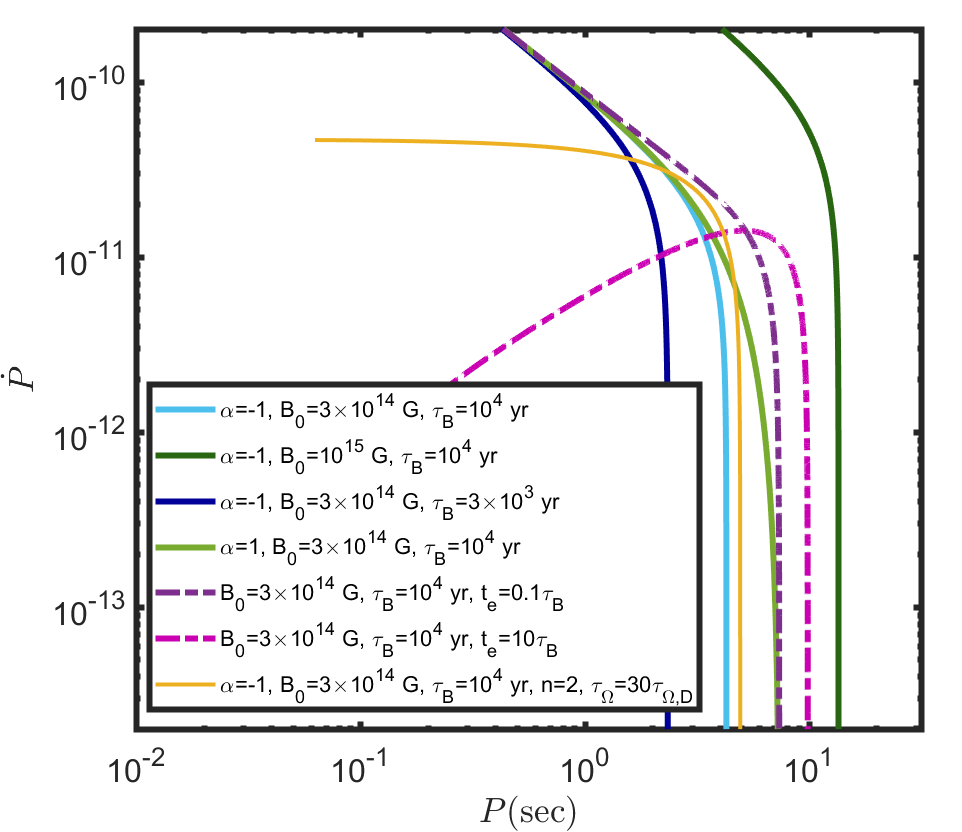}
		\caption{Evolutionary tracks in the $P-\dot{P}$ diagram for different model parameters (dashed lines represent magnetic field emergence models). In all cases, systems evolve from left to right over time.}
		\label{fig:ppdottrack}
	\end{figure}
	
	In Figure \ref{fig:ppdottrack} we explore the tracks followed by different evolution models in the $P-\dot{P}$ diagram. A given system evolves with time from left to right in this diagram.
	This figure demonstrates the evolutionary effects of the different model parameters:
	\begin{itemize}
		\item $B_0$ affects the `normalization' of the plot: larger values of $B_0$ result in parallel tracks to the right of the original track in the $P-\dot{P}$ diagram.
		\item $\tau_B$ affects the pivot point in the diagram, beyond which $\dot{P}$ starts decreasing more rapidly: lower values of $\tau_B$ push this point to the left of the diagram.
		\item $\alpha$ affects the steepness of $\dot{P}$'s evolution as a function of $P$ beyond the pivot point. Larger values lead to shallower evolution.
		\item $n$ affects mainly the evolution before the pivot point. Lower values lead to a shallower decrease of $\dot{P}$ as a function of $P$. For a constant field $\dot{P}\propto P^{n-2}$. $n=2$ represents a critical value, below which $\dot{P}$ increases with $P$ (prior to the pivot point).
		\item $t_e$ provides another way to start with a small value of $\dot{P}$. In models with $t_e>\tau_B$, $\dot{P}$ increases with time, until $t_e$ and only then starts decreasing; larger $t_e/\tau_B$ leads to a lower initial value of $\dot{P}$ at a given $P$.
	\end{itemize}
	Because the different model parameters affect the distribution of $P$ and $\dot{P}$ in unique ways, there are no strong degeneracies between these parameters. This renders them easier to determine by virtue of the likelihood analysis in \S \ref{sec:like}.

\begin{figure*}
		\centering
		\includegraphics[width=0.48\textwidth]{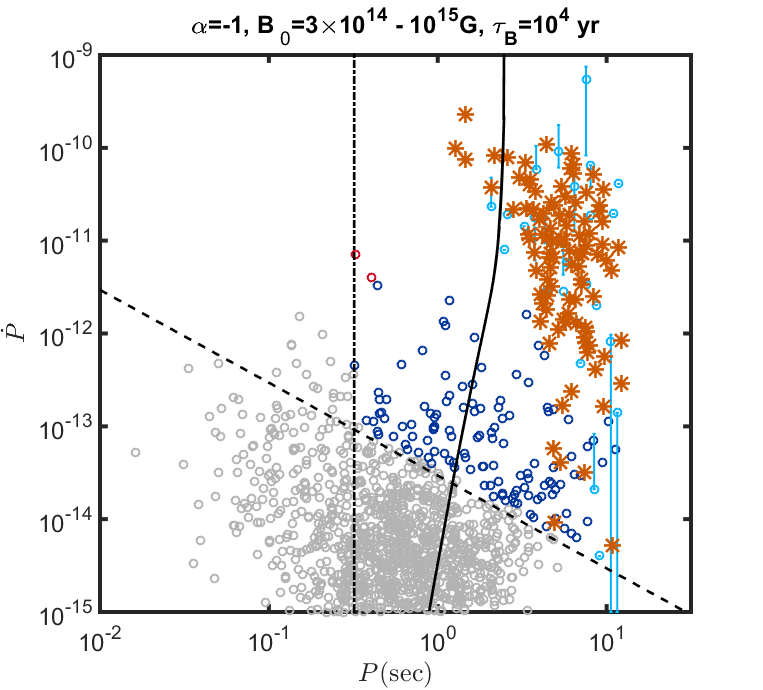}
		\includegraphics[width=0.48\textwidth]{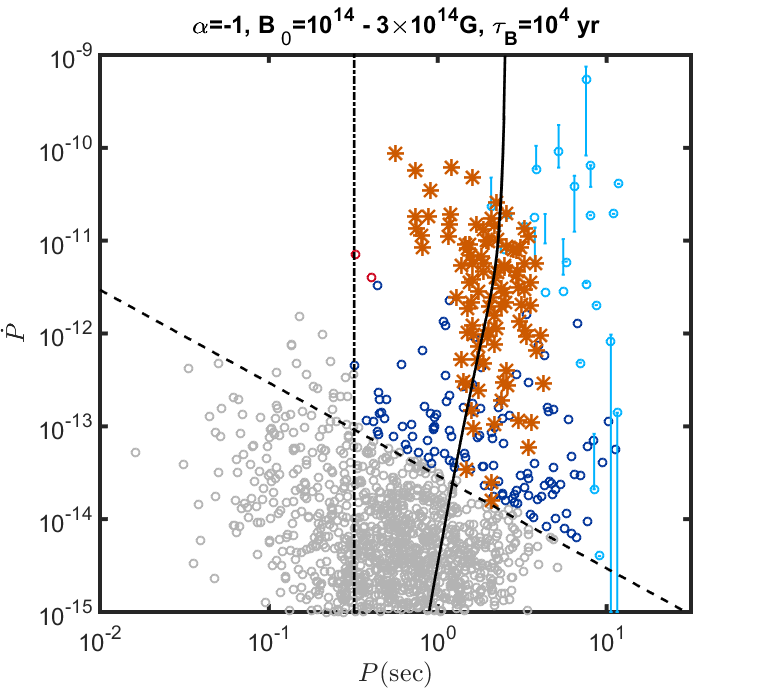}\\
		\includegraphics[width=0.48\textwidth]{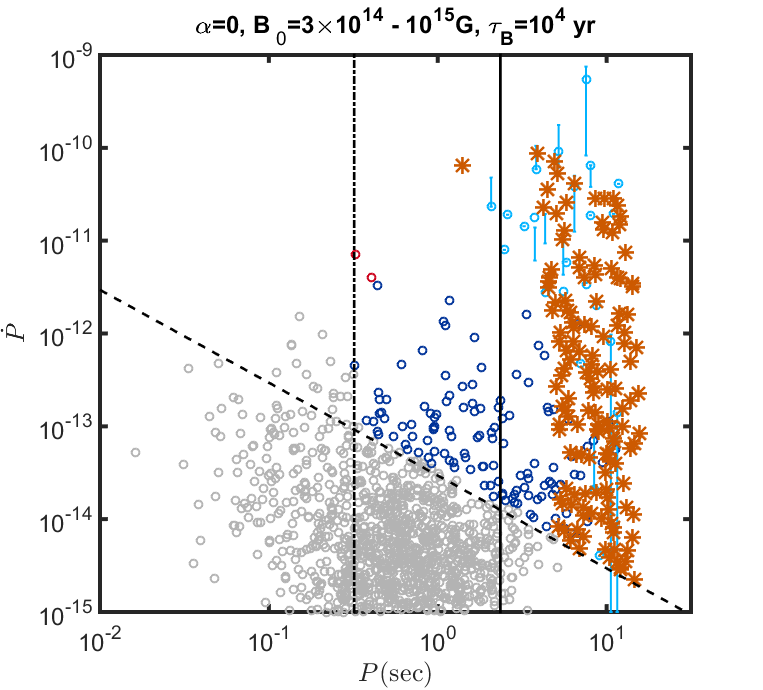}
		\includegraphics[width=0.48\textwidth]{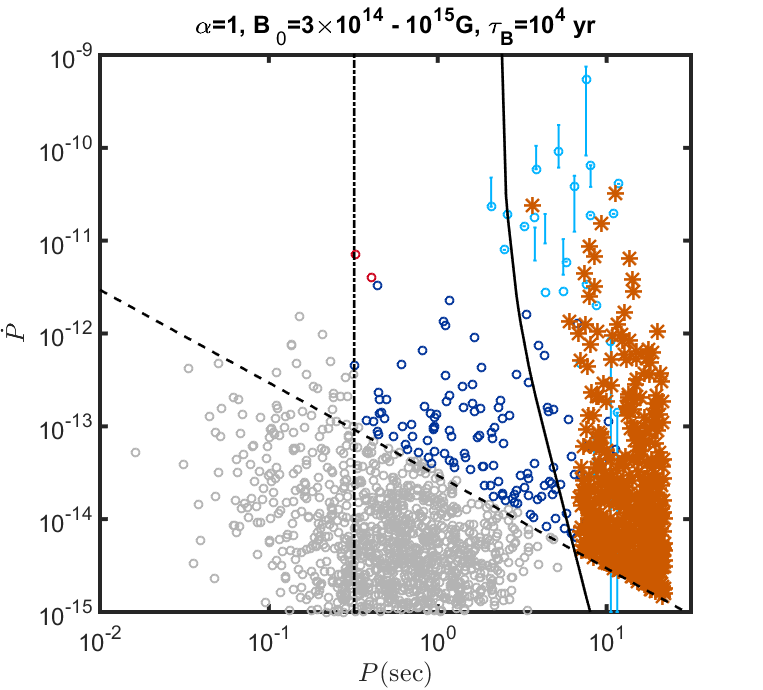}\\
		\includegraphics[width=0.48\textwidth]{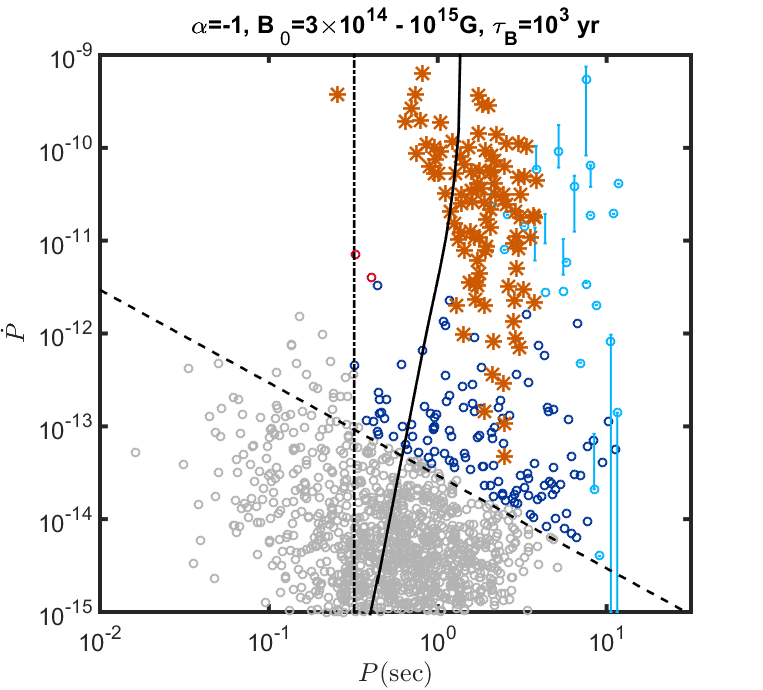}
		\includegraphics[width=0.48\textwidth]{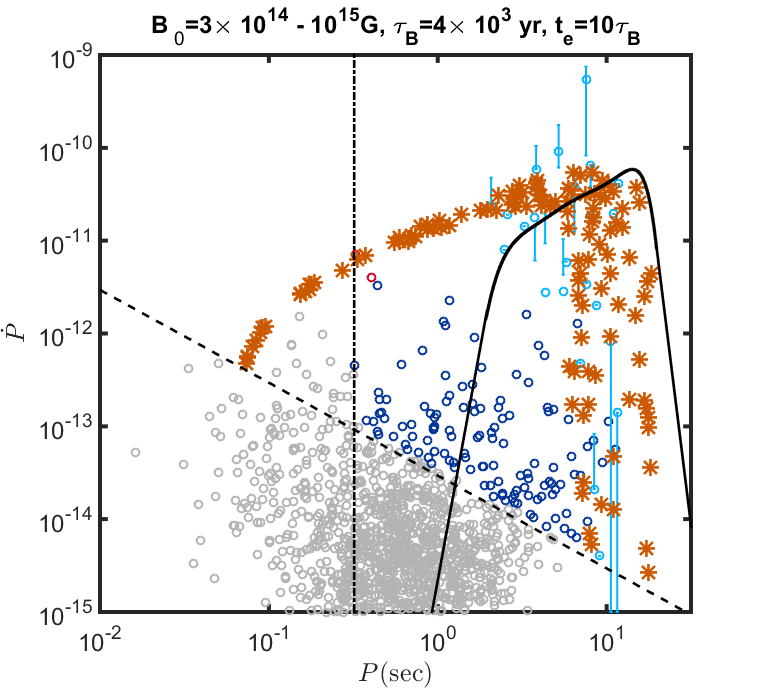}\\
		\caption{Monte Carlo simulations of the observable magnetar population for different model parameters. The simulated magnetars are shown in orange asterisks, while the observed populations of magnetars are denoted by the same symbols as in Figure~\ref{fig:ppdot}. Solid black lines denote the critical line, to the right of which magnetic energy losses become stronger than rotational energy losses.}
		\label{fig:ppdotmodels}
	\end{figure*}

	To fully visualize the effects of these parameters on the magnetar distribution, we plot in Figure~\ref{fig:ppdotmodels} representative realizations of a Monte Carlo simulation in which the birth times of different systems are taken to be uniformly distributed within the last ten Myr \footnote{The time-scale is chosen such that it is longer then it takes the magnetic field to decay below that of the weakest field magnetar. This ensures that within the regime of interest, a steady state has been established.}, and the initial magnetic fields are log-uniformly distributed between $B_{0,\rm min}-B_{0,\rm max}$. We can draw the following conclusions from these plots:
	\begin{itemize}
		\item $B_0$ needs to span roughly half an order of magnitude (one and a half orders of magnitude) in order to reproduce the full range of confirmed magnetars (magnetar candidates) $P$ and $\dot{P}$ values. A typical range that matches the observations of the former group is $B_0=3\times 10^{14}-10^{15}G$.
		\item Models with $\alpha\gtrsim 0$ are disfavored by the data of confirmed magnetars alone, since they lead to a `bottom heavy' distribution of magnetars in the $P-\dot{P}$ diagram, contrary to observations of those objects (sample A) which cluster at larger values of $\dot{P}$. It is, however, unclear if this is due to a selection effect, since the magnetars at lower $\dot{P}$ have intrinsically lower luminosities and may be more easily missed (see discussion in \S \ref{sec:lum} and \S \ref{sec:logNlogS}). In addition, models with $\alpha>0$ naturally lead to a region in which $\dot{E}_B>\dot{E}_{\Omega}$ that becomes narrower for larger periods, as is the case in the observed sample of confirmed magnetars. For these reasons we explore in this work also scenarios with larger values of~$\alpha$.   
		\item Models with magnetic field emergence, such that $t_e>\tau_B$, provide an intriguing explanation to the magnetar candidates, PSR J1119-6127 and PSR J1846-0258, which is that these are `magnetars in the making'. This also explains the young magnetic ages inferred for those systems (see Table~\ref{tbl:ages}). These models, however, struggle to explain the general observed population. They predict a significant number of magnetars between the two magnetar candidates and the confirmed magnetar population, and to the bottom left of the two magnetar candidates, neither of which are observed. There is no clear observational bias against detections in both areas, as these should contain some of the brightest systems in both radio and X-rays. Indeed, the former region is significantly devoid of {\it any} neutron stars, suggesting that confirmed magnetars represent a distinct population of neutron stars.		
	\end{itemize}
	
	\section{Likelihood analysis}
	\label{sec:like}
	Using the general spin and magnetic field evolution models discussed above, our next step is to determine the conditions necessary to fit the observed spin periods and period derivatives of the magnetar population.
	
	\subsection{Statistical model}
	We use a maximum likelihood method to find the most likely physical conditions that reproduce the magnetar population. The evolution of the magnetic field requires three parameters: the typical magnetic field at birth, $B_{0,\rm min}$; the magnetic field decay time, $\tau_B$; and a final parameter describing the functional form of the evolution, given by $\alpha$ for the case of field decay, or by the emergence time $t_e$ in the case of magnetic field emergence. For a dipole field, the evolution of the spin requires one additional parameter, $\Omega_0$, the typical spin frequency at birth. We stress that as long as $\Omega_0\gg\Omega$, this parameter has a very small effect on shaping the distribution. We fix $\Omega_0=100\mbox{ s}^{-1}$, but have verified that taking values in excess of $100\mbox{ s}^{-1}$ does not affect our results.
	
	The likelihood function is given by:
	\begin{eqnarray}
	\label{eq:Likelihood}
	&  L(B_0,\tau_B,\alpha,\Omega_0)=\Pi_i P(B_0|B_{0,\rm min}) P(t_{\rm form})\nonumber \\& \times P(\Omega_i,\dot{\Omega}_i|B_0,t_{\rm form},\tau_B,\alpha,\Omega_0)
	\end{eqnarray}
	where $i$ runs over the observed systems in the sample, $P(B_0|B_{0,\rm min})$ is the probability of a magnetic field $B_0$ at birth given a log-uniform distribution between $B_{0,\rm min}-B_{0,\rm max}$ (with $B_{0,\rm max}=3 B_{0,\rm min}$ for sample~A and $B_{0,\rm max}=30 B_{0,\rm min}$ for sample~B; see \S \ref{sec:model}), and $P(t_{\rm form})$ is the probability of a given magnetar formation time which depends on the star formation rate in the Galaxy. Since on the time-scales of $\lesssim 10\mbox{ Myr}$, which are relevant to the magnetar population, the latter is roughly constant, we assume a uniform distribution between 0 and 10~Myr ago. Finally, $P(\Omega_i,\dot{\Omega}_i|B_0,t_{\rm form},\tau_B,\alpha,\Omega_0)$ is the conditional probability of obtaining the observed spin-frequency and spin-frequency derivative of neutron star $i$ given the model parameters and probabilities above. For the model involving magnetic field emergence, the term $\alpha$ in Equation~\ref{eq:Likelihood} is replaced with $t_e$. Finally, deviating from the dipole case requires adding two more parameters: $n$ and $\tau_{\Omega}$.
	
	Taking $\Omega_0$ fixed as described above leaves us with three free parameters in the dipole models and five in the non-dipole models.  We search for the model parameters that maximize the likelihood function above. For any given model, the likelihood is calculated by means of a Monte Carlo simulation. We draw $10^6$ systems according to the probabilities above, evolve them in time until the present day, and check which of those systems would currently be observed as magnetars. This is done by computing the `dipole-equivalent' magnetic field (i.e. the field determined by $P,\dot{P}$ using the dipole formula) and requiring that it satisfies the condition $B_{\rm dip}>6\times 10^{12}$~G. Out of those systems, we find for each observed magnetar $i$ the fraction of simulated magnetars that are consistent with $P_i$ to within a factor of 2 and $\dot{P}_i$ to within a factor of 3. We allow for a slightly wider factor of $\dot{P}_i$ values than for $P_i$ to represent the fact that the former is observed to fluctuate more over time (and involves larger errors). The product of the probabilities of reproducing individual systems then constitutes the overall likelihood of a given model.
	
	\subsection{Likelihood results}
	\label{sec:likeres}
	The best fit parameters for our likelihood analysis are summarized in Table~\ref{tbl:model}, for the different samples and models described in \S \ref{sec:sample} and \S \ref{sec:model}. We focus here mainly on the sample of confirmed magnetars, sample A, and discuss how the modelling changes for sample B in \S \ref{sec:magcand}.
	The overall best model for sample A is given by the following parameters: $B_0=3\times 10^{14}- 10^{15}G, \tau_B=10^4\mbox{yr}, \mbox{ and } \alpha=-0.9$.
	These model parameters are consistent with our expectations based on the analysis outlined in \S \ref{sec:model}. Furthermore, this typical magnetic decay time is consistent with the conclusions of previous authors \citep{Kouveliotou1998,Colpi2000,Dall'Osso2012,Vigano2013}.
	
	\begin{table*}
		% 		\footnotesizea
		\begin{center}
			\caption{Best fit model parameters for magnetar population likelihood analysis. Quoted errors denote $2\sigma$ values.}
			\begin{threeparttable}
				\begin{tabular}{ccccc}\hline	
					Model &  $B_{0,\rm min}[10^{14}G]$ & $\tau_{B}\mbox{ [kyr]}$ & $\alpha$ & $t_e/\tau_B $ \\ \hline%& $n$ & $\tau_{\Omega}/\tau_{\Omega_0}$ \\ \hline
					Sample A: magnetic decay & $3_{-0.5}^{+2}$  & $10_{-7.5}^{+6}$ & $-0.9\pm0.2$ & -\\% & 1 & 1\\
					Sample A: magnetic emergence & $2.5_{-0.5}^{+1.5}$  & $4_{-1.5}^{+3}$ & - & $0.3_{-0.3}^{+0.3}$\\% & 1 & 1\\
					Sample B: magnetic decay & $0.5_{-0.1}^{+0.2}$  & $0.4_{-0.2}^{+0.2}$ & $0.7\pm0.1$ & -\\% & 1 & 1\\
					Sample B: magnetic emergence & $0.5_{-0.1}^{+0.2}$  & $1.6_{-0.6}^{+1}$ & - & $0.03_{-0.03}^{+0.03}$\\% & 1 & 1\\
					\hline  
					\label{tbl:model}
				\end{tabular}
			\end{threeparttable}
		\end{center}
	\end{table*} 
	
\begin{figure}
		\centering
		\includegraphics[width=0.35\textwidth]{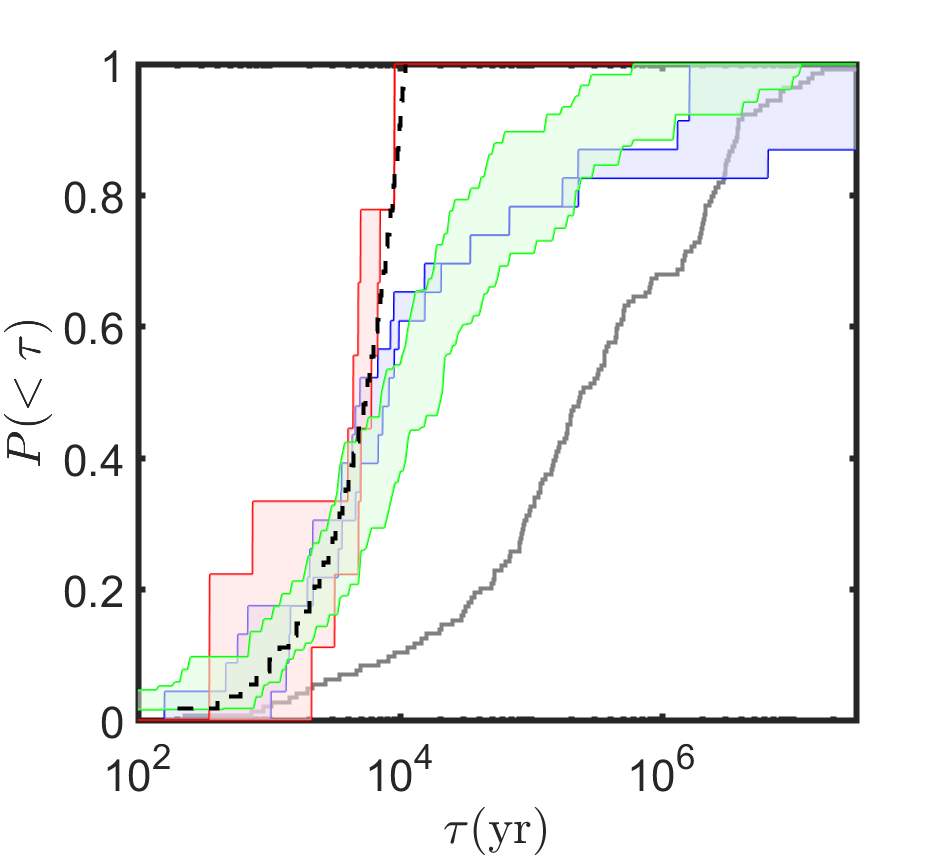}
		\caption{Cumulative distribution of magnetar age limits. The blue (green) region denotes the range of $\tau_{\Omega}$ values for the observed (simulated) population. The solid gray line depicts the same distribution for the entire population of magnetar candidates (sample B). These are upper limits on the real ages of the systems. The dashed line represents the real ages of the simulated systems for a typical realization, and the red region denotes the range of $\tau_{\rm SNR}$ values for the observed magnetars.}
		\label{fig:taudist}
	\end{figure}

	The fact that this model maximizes the likelihood function does not yet ensure that it provides a statistically adequate description of the data. To test this, we perform a Kolmogorov-Smirnov test between the observed and simulated distributions of $P$ and $\dot{P}$. The model is found to be statistically consistent with both of these distributions. In particular, this agreement can be shown via the distribution of $\tau_{\Omega}=|P/2\dot{P}|$ between the observed and simulated systems; see Figure~\ref{fig:taudist}. Indeed the two distributions, match very well. As a comparison, we also depict the same distribution for sample B (which significantly differs from the former two). We also plot the real ages of the simulated magnetars (which, as discussed in \S \ref{sec:rates}, are always equal or younger than the spin-down ages) and the age estimates of confirmed magnetars from SNR associations in the same figure. 
	Even though the statistical model does not use the SNR ages as input, the real ages implied by the model are consistent with the observed SNR ages, leading additional support to the modeling. Finally, we note that, as expected, at times shorter than $\tau_B$ there is a good agreement between the real ages and $\tau_{\Omega}$, while at longer times, $\tau_{\Omega}$ is significantly larger than the real ages. 
	The parameters of this model imply a Galactic magnetar birth rate of $\Phi_{\rm mag}=3.5_{-1.5}^{+3.5}\mbox{ kyr}^{-1}$ (at a 2$\sigma$ confidence level). This is consistent with the limits derived in \S \ref{sec:rates}.
	
	\subsection{Luminosity function}
	\label{sec:lum}
	For a constant rate of magnetar formation $dN/dt=\Phi_{\rm mag}$, the persistent luminosity function of magnetars (i.e., the number of objects with a given luminosity, assuming it to be powered by the magnetic energy losses) can be calculated analytically. Since the magnetar's field drops significantly after $\tau_B$ (see Figure~\ref{fig:Omegat}), we focus on times $t<\tau_B$ during which all systems born as magnetars are still observable as such. We derive
	\begin{equation}
	\frac{dN}{d\dot{E}_B}=\frac{dN}{dt}\frac{dt}{d\dot{E}_B}\propto {\dot{E}_B}^{-(2\alpha+2) \over \alpha+2} \ \mbox{ for } \ \dot{E}_B\lesssim \frac{B_0^2R^3}{6\tau_B}
	\end{equation}
	while at higher $\dot{E}_B$, the number of magnetars drops roughly exponentially.
	
\begin{figure*}
		\centering
		\includegraphics[width=0.39\textwidth]{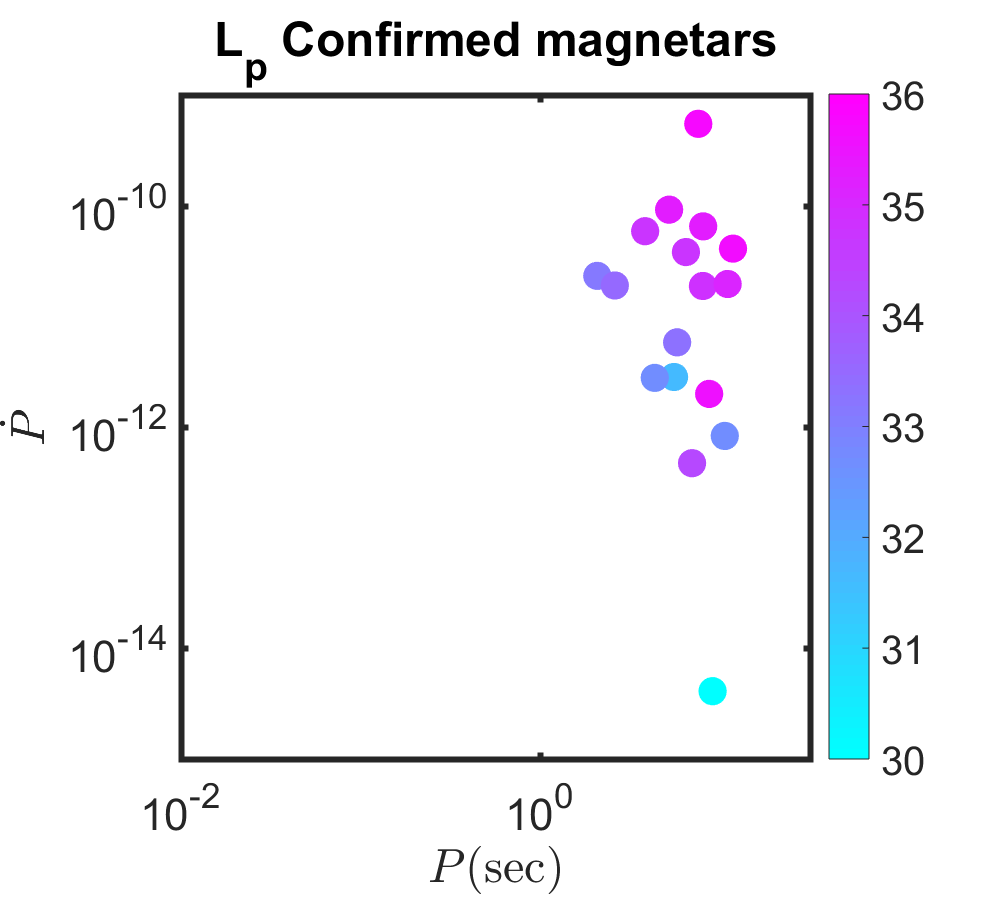}
		\includegraphics[width=0.39\textwidth]{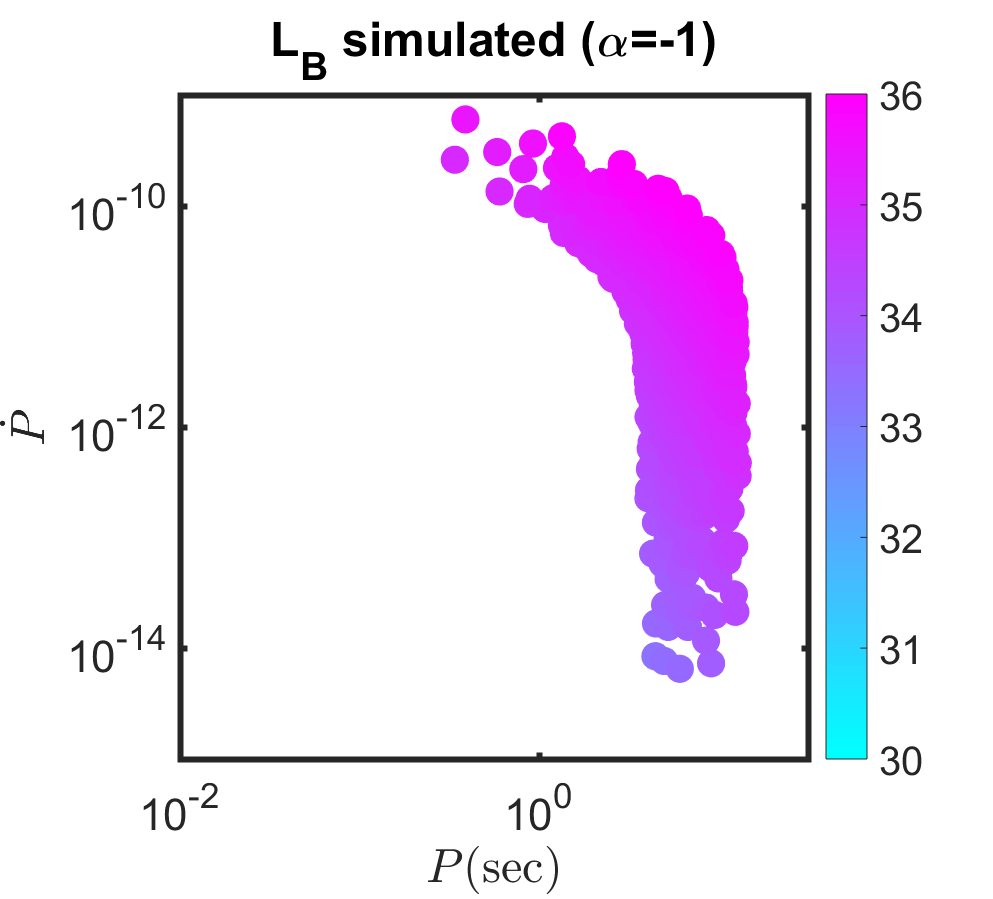}\\
		\includegraphics[width=0.39\textwidth]{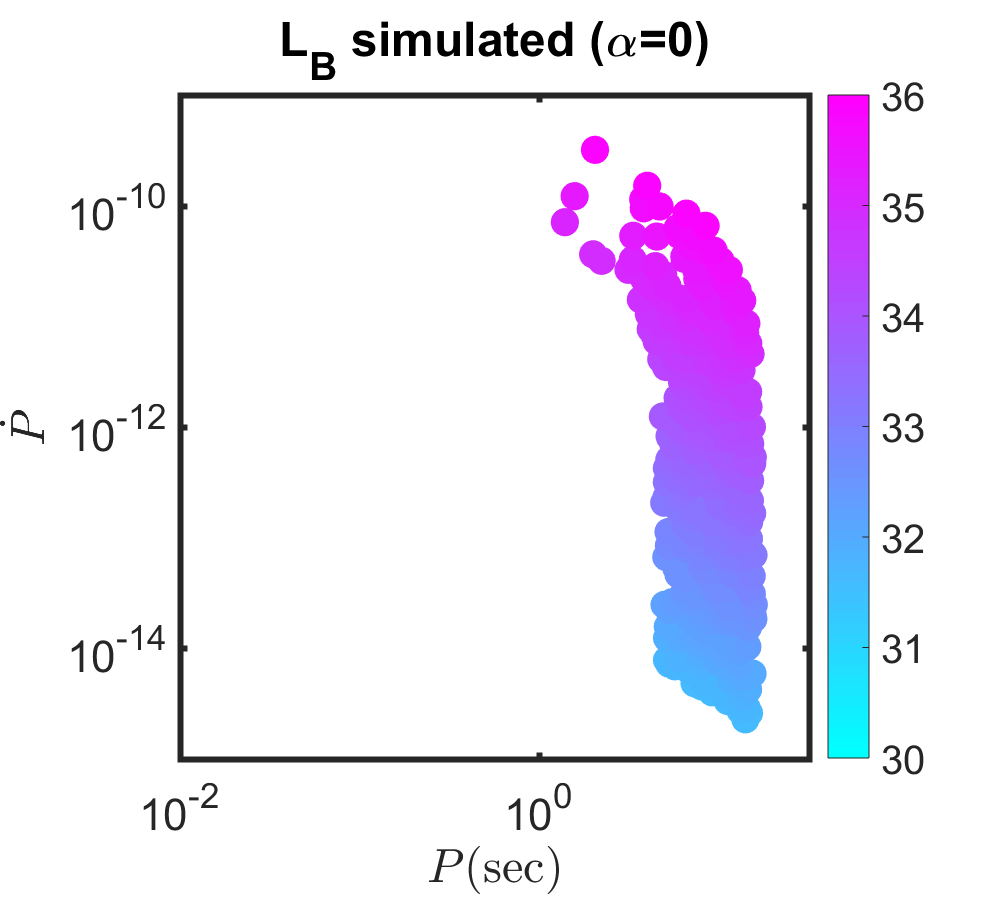}
		\includegraphics[width=0.39\textwidth]{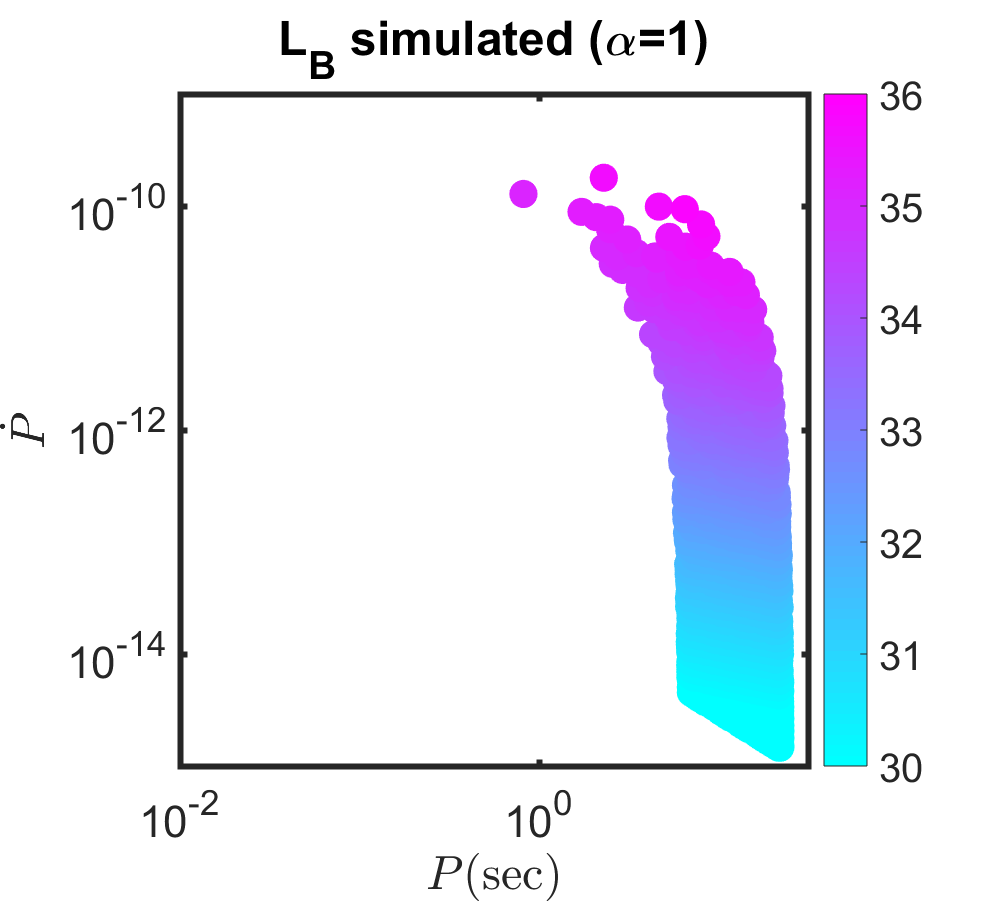}
		\caption{Top Left: persistent luminosities of confirmed magnetars in the $P-\dot{P}$ diagram. Other panels: magnetic energy loss rates for simulated magnetars assuming our best fit model for the magnetar population with $B_0=3\times 10^{14}-10^{15}G, \tau_B=10^4\mbox{yr}$, and different values of $\alpha$.}
		\label{fig:ppdotLp}
	\end{figure*}

	In our modelling we have not assumed any apriori relation between the persistent luminosity and the magnetic energy loss rate. Since the latter is the most likely energy source for the former, we expect to find $L_p<\dot{E}_B$ (recall that some of the magnetic energy lost is needed to power bursts, outbursts, and GFs). It is striking that for our best fitting model this condition is indeed satisfied, and that the two values exhibit similar trends in the $P-\dot{P}$ diagram (see Figure~\ref{fig:ppdotLp}), with an average
	value of $\langle\log_{10}(\dot{E}_B/L_p)\rangle=0.9$ for $\alpha=-1$, and an average of $\langle\log_{10}(\dot{E}_B/L_p)\rangle=0.6$ for $\alpha=0$.
	This lends strong support to the idea that the evolution of magnetars in the diagram is dominated by magnetic energy losses.

	\subsection{Magnetar Outbursts}
	\label{sec:burst}
\begin{figure}
		\centering
		\includegraphics[width=0.4\textwidth]{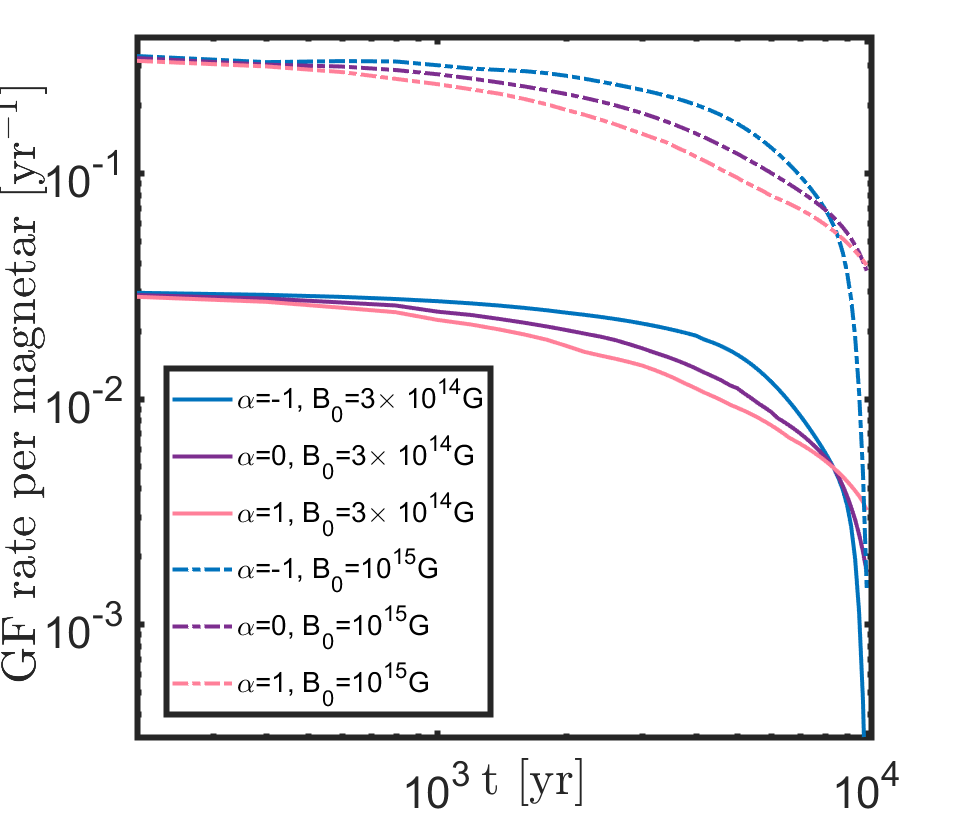}\\
		\includegraphics[width=0.4\textwidth]{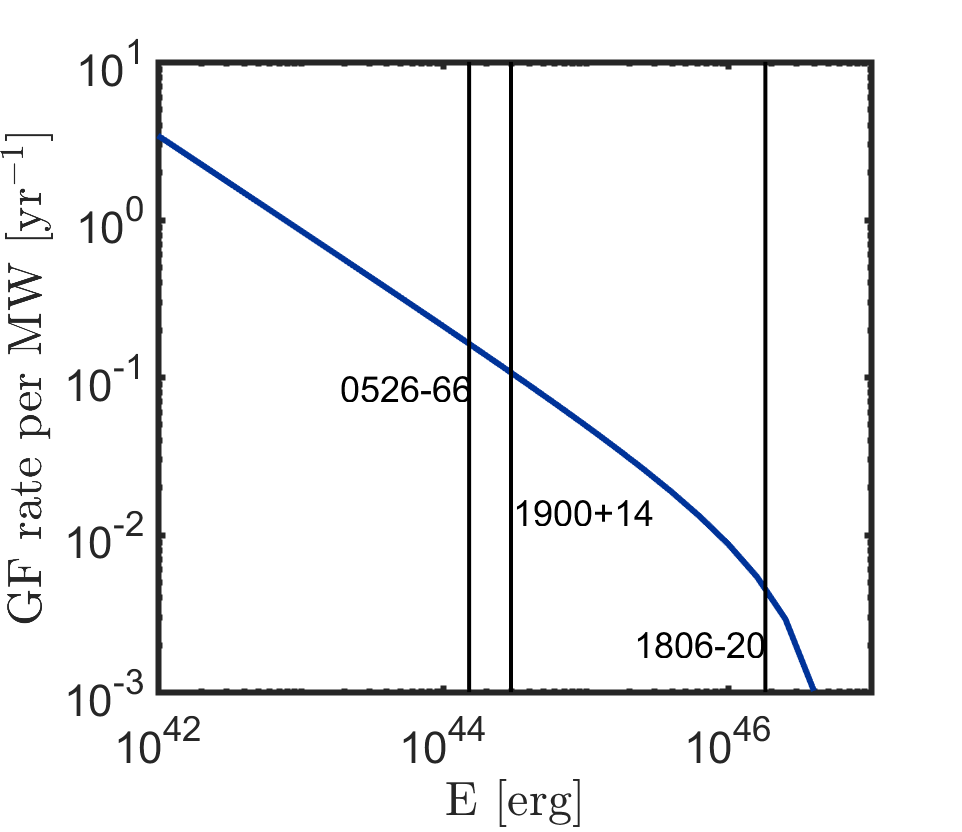}
		\caption{Top: Rate of GFs from a single magnetar, assuming a typical energy of $10^{44}$ erg and that most of the magnetic energy losses can be used to power the flares. We take here $\tau_B=10^4$ yr and different values of $\alpha$ and $B_0$. Bottom: Rate of flares with energy smaller than $E$ for the simulated population, based on our likelihood analysis of the confirmed magnetar population. Also shown are the energies of the three observed GFs (from \citealt{Tanaka2007}).}
		\label{fig:GF}
	\end{figure}

	The magnetic energy that is not dissipated through the persistent luminosity, must still be dissipated from the neutron star by some other means. The most likely scenario is that it is used to power the bursting activity of magnetars, which occur on a large range of time scales and energies. With an observed energy distribution $dN/dE\propto E^{-\gamma}$ with $\gamma=1.4-1.8$ \citep{Gogus1999,Gogus2000} during outbursts, the total energy release $\propto E^2dN/dE$ is likely dominated by the most energetic events. Therefore, we assume here that the GFs dominate the energy release and calculate their required rate in order to release all of the magnetic energy. We further assume that the typical energy of a GF is $10^{44}$ erg and that their rate scales with the overall magnetic energy loss rate $\dot{E}_B/E_{B,0}$. Under this assumption we calculate the rate of the events for different values of $\alpha$ and $B_0$ and for a single magnetar. The results are depicted in the top panel of Figure~\ref{fig:GF}. For a magnetar with an initial field of $3\times 10^{14}$~G ($10^{15}$~G), at a typical age of $5\times 10^3$ yr, GFs are expected to occur at a rate $\sim 10\mbox{ kyr}^{-1}$ ($\sim 100\mbox{ kyr}^{-1}$). This is consistent with the observed rate of GFs (see \S \ref{sec:rates}).

	The next step is to calculate the rate of flares for the entire magnetar population. We assume here that the observed burst distribution with $dN/dE\propto E^{-\gamma}$, for $E>10^{36}$~erg and $\gamma=1.6$, can be extended up to the GFs. We take the maximum for the distribution to be the total magnetic energy of the magnetar. We further assume that the rate of flares is proportional to $\dot{E}_B$. Using our best fit models from the likelihood analysis, we generate a simulated population of magnetars, and calculate the rate of flares with energy larger than $E$. The rates are almost entirely independent of $\alpha$, as changing $\alpha$ affects mostly the population of older magnetars, which contribute very little to the overall energy released in flares. For this reason we show only the single value of $\alpha=-1$ in the bottom panel of Figure~\ref{fig:GF}. We find that approximately one GF with $E>10^{44}$ erg ($E>10^{46}$ erg) should occur every $6$ years ($200$ years).
	
	\subsection{log$N$-log$S$}
	\label{sec:logNlogS}
	To estimate the number of faint magnetars that have not been detected, we compare the expected log$N$-log$S$ distribution of persistent fluxes with the observed one. Here we use the observed log$N$-log$S$ distributions with both absorbed and unabsorbed fluxes listed in \cite{Olausen2014}. We calculate log$N$-log$S$ assuming that 
	the spatial density of Galactic magnetars is proportional to the stellar density of the thin disk \citep{McMillan}. For the distribution with absorbed fluxes, we assume that the X-ray absorption cross section per hydrogen for $2$--$10$ keV \citep{Wilms2000} and the interstellar medium density are both uniform in the Galactic thin disk, so that the optical depth is given by
	\begin{eqnarray}
	\tau(D) = \sigma n_H D \approx \tau_0 \left(\frac{D}{1\,{\rm kpc}} \right),
	\end{eqnarray}
	where $D$ is the distance from Earth and $\tau_0$ is a constant,
	e.g., $\tau_0\approx 0.07$--$0.15$ for $n_H=1\,{\rm cm^{-3}}$ at $2$ keV.
	We count the number of magnetars with an X-ray flux brighter than $S_{\rm unabs}=L_p/4\pi D^2$ or $S_{\rm abs}=L_pe^{-\tau(D)}/4\pi D^2$ for a given magnetar luminosity function. 
	
\begin{figure*}
		\centering
		\includegraphics[width=0.39\textwidth]{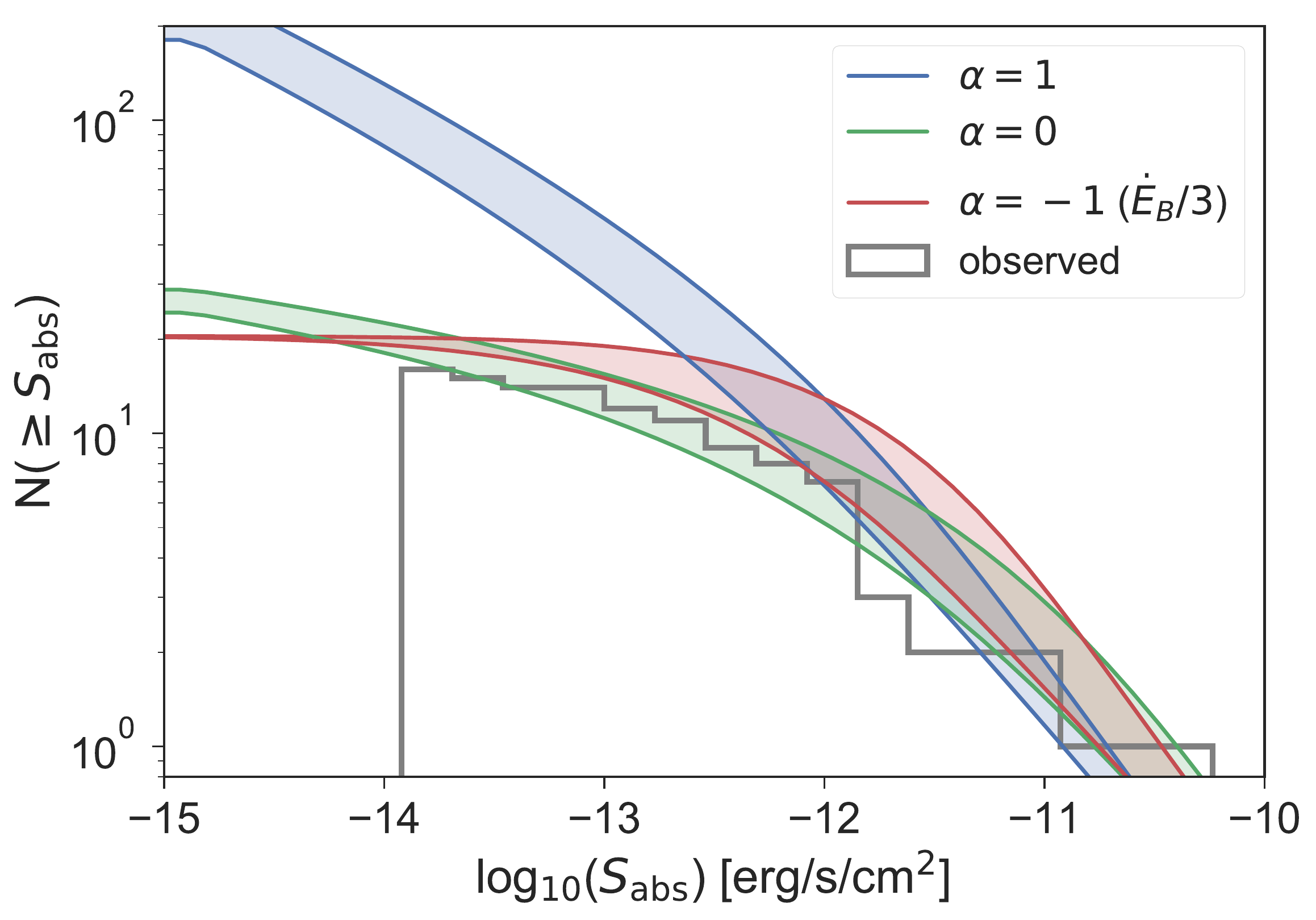}
		\includegraphics[width=0.39\textwidth]{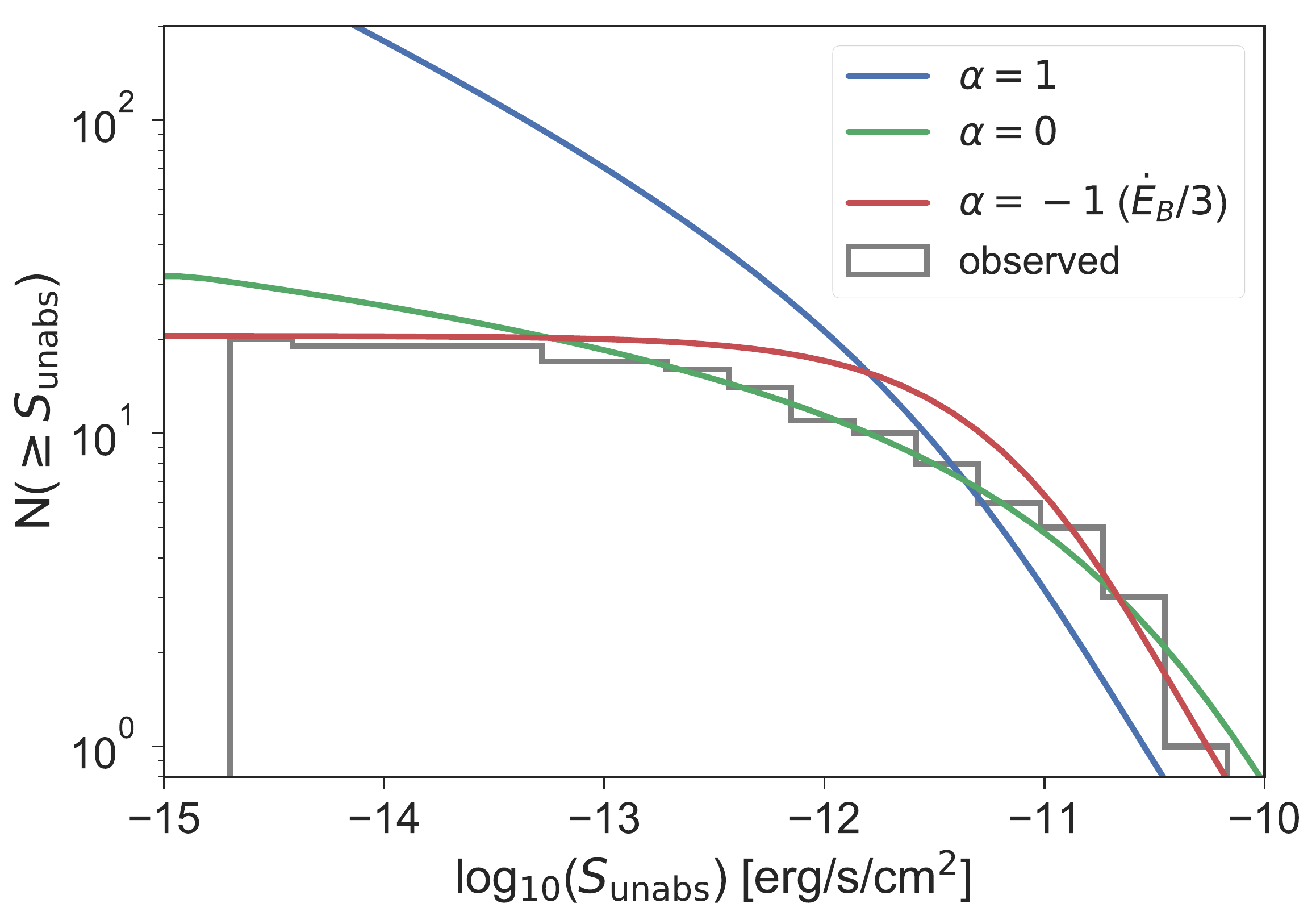}
		\caption{Log$N$-log$S$ plot for different magnetic field evolution models, and for absorbed and unabsorbed X-ray fluxes in the left and right panel, respectively. We assume that the persistent X-ray luminosity $L_p$ is approximately $\dot{E}_B$, but $L_p \approx \dot{E}_B/3$ is assumed for $\alpha=-1$. To calculate the absorbed fluxes, we take the optical depth parameter $\tau_0$ to be $0.1$--$0.25$. }
		\label{fig:logN}
	\end{figure*}

	Figure \ref{fig:logN} compares the calculated log$N$-log$S$ for $\alpha=-1,\,0,\,1$ with the observed ones. Here we assume that  the persistent X-ray luminosity $L_p$ is approximately $\dot{E}_B$, but $L_p \approx \dot{E}_B/3$ is assumed for $\alpha=-1$. To calculate the absorbed fluxes, we take the optical depth parameter $\tau_0$ to be $0.1$--$0.25$.
	
	The observed log$N$-log$S$ is reasonably reproduced for $\alpha=0$
	and $-1$. The match between the observed distribution and our best fit model $\alpha\approx -1$ is striking. We stress that this match is independent of our likelihood fitting procedure, which was only based on the distribution of $P$ and $\dot{P}$ of the observed magnetars, and was not informed by the distribution of observed fluxes. We further note that models with $\alpha\approx -1$ and $\alpha=0$ predict that the fraction of Galactic magnetars missing from current observations is moderate at best, between $\sim0-0.3$. We note however that the observational bias in detecting magnetars is highly uncertain due to different systems being detected in different modes (bursting versus persistent) and with different missions. Because of this reason it is not trivial to determine a threshold flux limit above which the sample can be considered to be complete. Indeed, the lowest magnetic field magnetar, SGR 0418+5729, is also one of the closest to Earth. Its intrinsic persistent luminosity is estimated to be $\sim 9\times 10^{29}\mbox{erg s}^{-1}$, more than four orders of magnitude below the median of the magnetar population. If many more such systems exist in the Galaxy but are missed, since they are simply too dim to be observed, the magnetar $P-\dot{P}$ distribution would become more `bottom heavy'; and as a result, the inferred value of $\alpha$ would increase. An intrinsic value of $\alpha=1$ would imply that the fraction of missing magnetars is $\lesssim 10$. A systematic survey of some portion of the Galaxy in search for magnetars would be ideal for distinguishing between these possibilities. We have recently started this project with the Neil Gehrels {\it Swift} Observatory.
	
	\subsection{Dependence on model parameters}
	\label{sec:likedepend}
	The results for the model with magnetic field emergence are shown in Table~\ref{tbl:model}. As explained in \S \ref{sec:model}, models with $t_e\gtrsim \tau_B$ provide a very poor description of the data. In the $P-\dot{P}$ diagram there are no pulsars showing magnetar-like behaviour in between PSR J1119-6127, PSR J1846-0258 and the confirmed magnetars, or to the left of the first two. This implies that field emergence in the form given by Equation~\ref{eq:revive} is unlikely to play a significant role in the evolution of most magnetars. More complicated models of field emergence may not be ruled out, but without the specific theoretical motivation, they would introduce too many free parameters to the statistical modeling. For the same reason, models with $n\lesssim 2$ are strongly disfavored by the data. Furthermore, models with $n\neq 3$ add two free parameters to the model (see \S \ref{sec:model}), without improving the overall likelihood, and are therefore not required by the data.
	
	\subsection{Comparison of magnetar candidates to confirmed magnetars}
	\label{sec:magcand}
	If the magnetar candidates PSR J1119-6127 and PSR J1846-0258 are to be considered as part of the magnetar population, the estimated magnetar formation rates (see Equation~\ref{eq:rates}) would increase by a factor of $\sim 3-5$ and become a sizable fraction of the overall pulsar or massive star formation rate. This suggests that these systems represent an evolutionary path distinct from that of the confirmed magnetars. Indeed, this is supported by the gap between the locations of those systems in the $P-\dot{P}$ diagram (see \S \ref{sec:model}) and those of confirmed magnetars. Interestingly, for both magnetar candidates, the SNR age is somewhat greater than the upper limit from spin-down. Furthermore, there is a much larger discrepancy between the magnetic age limit for those systems and the SNR age, by more than two orders of magnitude in one case. This suggests that the magnetic ages of those systems are unrealistic, which is possible in models where the surface magnetic field does not monotonically decline over time (see magnetic emergence model, \S \ref{sec:model}). Such a model, although disfavored by our likelihood analysis for the magnetar population as a whole (see \S \ref{sec:likedepend}), may still be applicable to those two systems. It would also imply that these systems will, in the future, move towards the bulk of the magnetar population, thus providing a natural explanation for their magnetar-like nature.
	
	An alternative possibility is to attempt to explain all pulsars with spin periods and magnetic fields at least as large as those of the confirmed magnetars and the magnetar candidates (i.e., sample B) with a single model. The results of this analysis, presented in Table~\ref{tbl:model}, are statistically consistent with the $P$ and $\dot{P}$ data. However, they do not provide a good understanding of the persistent X-ray luminosities, because they predict that all pulsars with large periods and small period derivatives should appear as magnetars, i.e., they should have a persistent luminosity that is larger than their energy loss from spin-down, see Figure~\ref{fig:ppdotmodels}, which is contrary to the observations.
	
	\section{Conclusions}
	\label{sec:conclusions}
	Current theory seems to favor rarity of the magnetar population. Various models point towards extreme conditions in magnetar progenitors, regarding the initial rotation and/or magnetic field, among other variables \citep{Maeder+Meynet2000,Heger+Langer2000,Hirschi+2004}. These theoretical considerations would naively imply a limited range of magnetar progenitors, specifically of the progenitor masses.  The actual rate of magnetar formation may therefore be used to gain insight on two of the most basic questions in the field: {\it how are magnetars formed?} and {\it what are the progenitors of magnetars?}. 
	
	In this work, we have studied the formation rates and evolutionary channels of magnetars that are consistent with the observed population. Using the measured spin-down rates, magnetic energy loss (as probed by X-ray emission), and association with SNRs, we find that $2.3-20$ magnetars are born in the Galaxy every kyr (quoted range is at a $2\sigma$ confidence level). Using the Galactic star formation rate, as well as the entire population of observed SNRs and pulsars, we can derive also the equivalent rates for the general neutron star population: $17\pm 2\mbox{ kyr}^{-1}$. We conclude that $0.4_{-0.28}^{+0.6}$ of neutron stars are born as magnetars. Such a high rate is challenging from a theoretical point of view, as discussed above, and may be used to further constrain stellar population synthesis models.
	
	Studying the population of Galactic magnetars in the $P-\dot{P}$ diagram, we constrain the required initial conditions and magnetic spin-down evolutions that are permitted by observations. The best fit model reproduces the observed population well. In addition, it reproduces the formation rates discussed above and naturally accounts for the persistent luminosities of magnetars. The latter are found to represent a fraction of $\gtrsim 0.15$ of the magnetic energy losses.
	
	A large fraction of the magnetic energy losses of magnetars are expected to power their energetic outbursts. We find that GFs with $E>10^{44}$ erg ($10^{46}$ erg) should occur in the Galaxy at a rate of $\sim 0.15\mbox{ yr}^{-1}$ ($\sim 0.005\mbox{ yr}^{-1}$). Assuming a GF with $E\geq 10^{46}$ erg can be detected up to a distance of $\sim 100$~Mpc with {\it Swift}, and taking a {\it Swift} detection efficiency of $\sim 0.1$ \citep{3rdBAT}, we predict that $\sim 5$ GFs with energies above $10^{46}$ erg should be detectable every year with {\it Swift}. Such events could be hidden within the {\it Swift} population of short GRB with unconfirmed redshift. Natural candidates would be short {\it Swift} GRBs with no afterglow detection (beyond several minutes) in X-rays or optical. Indeed, using the {\it Swift} database \footnote{\url{https://swift.gsfc.nasa.gov/archive/grb_table/}}, we find that $\sim 1/3$ of events with $T_{90}<2$ s (corresponding to 81 events in the last 14 years, consistent with the rates above) have no afterglow detection. This fraction increases as $T_{90}$ decreases (going up to $\sim 2/3$ for $T_{90}<0.1$ s). In contrast, only a fraction $0.15$ of bursts with $T_{90}>2$ have the same property. These numbers are in agreement with the level of contamination of the population due to GFs suggested here. However, these estimates are made without taking into account visibility constraints for satellite repointing.
	
	The initial magnetic field of the magnetars at birth can be constrained to the range $B_0=3\times10^{14}-10^{15}$G (see also \citealt{Pons2011,Gullon2015}). This is consistent with magnetic field values that are required for millisecond magnetars to power superluminous supernovae \citep[SLSNe;][]{Metzger2015,Metzger2018} and gamma-ray bursts \citep[GRBs;][]{Beniamini2017,Metzger2018}. The other important ingredient for powering those energetic explosions is the initial spin of the magnetar, $\Omega_0$. We have shown here that the latter is not well constrained by observations of the Galactic magnetars. Indeed, given the extreme rarity of SLSNe and GRBs, only a very small faction, $\sim 10^{-3}-10^{-2}$ of magnetars need to be born with very high spins (close to break-up) in order to power SLSNe and GRBs.
	
	To account for the observed magnetar population, the magnetic field of magnetars has to decay significantly on a timescale of $\tau_B\sim 10^4$yrs. This is consistent with previous findings for the magnetar population \citep{Kouveliotou1998,Colpi2000,Dall'Osso2012,Vigano2013}, and shorter than the decay time inferred from the general pulsar population \citep{Gonthier2004,Vigano2013} which can be as long as $3$~Myr.
	If the low number of magnetars with large $P$ and low $\dot{P}$ within the confirmed sample of magnetars is representative of the intrinsic population, the magnetic field should decline roughly as $B=B_0(1-t/\tau_B)$ (corresponding to $\alpha=-1$ with $\dot{B}\propto B^{1+\alpha}$). This would imply that the magnetar sample in the Galaxy is close to being complete with a missing fraction of only $\lesssim 0.3$. However, based on the persistent luminosity and $\log N-\log S$ distributions, it is possible that we are still missing many lower-luminosity magnetars. In this case, $\alpha$ and the fraction of missing magnetars may become larger. For instance, for $\alpha=1$ the missing magnetars could outweigh the observed ones by a factor of ten. The missing objects should have unabsorbed fluxes $\lesssim 10^{-13}\mbox{ erg cm}^{-2}\mbox{ s}^{-1}$.
	
	Finally, we note that magnetic field emergence models can provide an explanation for the magnetar candidates PSR J1119-6127 and PSR J1846-0258, as `magnetars in the making'. In this scenario, those systems will evolve with time towards larger $P$ and $\dot{P}$, consistent with the bulk of the magnetar population. This scenario, however, is statistically disfavoured as an explanation for the evolution history of the majority of the magnetar population. Indeed, the gap in the $P-\dot{P}$ diagram between the magnetar candidates and the rest of the magnetar population suggests that magnetars are a distinct population of neutron stars.
	\voffset = -0.7in  
	\section*{Acknowledgements}
	We thank Mathew Baring, Zorawar Wadiasingh, George Younes, Tsvi Piran, Todd Thompson, and Chris Fryer for helpful discussions.
	\voffset = -0.7in  
	%%%%%%%%%%%%%%%%%%%%%%%%%%%%%%%%%%%%%%%%%%%%%%%%%%
	
	%%%%%%%%%%%%%%%%%%%% REFERENCES %%%%%%%%%%%%%%%%%%
	
	% The best way to enter references is to use BibTeX:

	%%%%%%%%%%%%%%%%%%%%%%%%%%%%%%%%%%%%%%%%%%%%%%%%%%
	
	%%%%%%%%%%%%%%%%% APPENDICES %%%%%%%%%%%%%%%%%%%%%
	
	%\appendix
	
	%\section{Some extra material}
	
	%%%%%%%%%%%%%%%%%%%%%%%%%%%%%%%%%%%%%%%%%%%%%%%%%%

	% Don't change these lines
	\bsp	% typesetting comment
	\label{lastpage}
\end{document}